\documentclass[12pt,draftcls,peerreviewca,onecolumn,a4paper]{IEEEtran}
\usepackage{amsfonts}
\usepackage{slashbox}
\usepackage{amsmath}
\usepackage{stmaryrd}
\usepackage{multirow}
\usepackage{graphicx}
\usepackage{amsfonts}
\usepackage{amsmath,epsfig}
\usepackage{mathbbold}
\usepackage{stfloats}
\usepackage{amssymb}
\usepackage{cite}
\usepackage{amssymb}
\usepackage{latexsym}
\usepackage{epsfig}
\usepackage{mathrsfs}
\usepackage{algorithmic}
\usepackage{algorithm}
\usepackage{subfigure}
\usepackage{slashbox}
\usepackage[all]{xy}
\usepackage{pifont}
\usepackage{bbding}
\usepackage{stmaryrd}
\usepackage{amssymb}
\usepackage{amsfonts}
\usepackage{epic}
\usepackage{stfloats}
\usepackage{latexsym}
\usepackage{epic}
\usepackage{bm}
\usepackage{xcolor}
\usepackage{mathrsfs}
\usepackage{pifont}
\usepackage{bbding}
\usepackage{amsmath,epsfig}
\usepackage{mathbbold}
\usepackage{epic}
\usepackage{enumerate}
\usepackage{stfloats}
\usepackage{latexsym}
\usepackage{epstopdf}
\usepackage{epic}
\usepackage{multirow}
\usepackage{stfloats}
\usepackage{bm}
\usepackage{booktabs}
\usepackage{color}
\usepackage{stmaryrd}
\usepackage{amsfonts}
\usepackage{amssymb}
\usepackage{stfloats}
\usepackage{cite}
\usepackage{graphicx}
\usepackage{psfrag}
\usepackage{subfigure}
\usepackage{array}
\usepackage{eurosym}
\usepackage{setspace}
\begin{document}

\title{Fundamental Green Tradeoffs: Progresses, Challenges, and Impacts on 5G Networks}

\newtheorem{Thm}{Theorem}
\newtheorem{Lem}{Lemma}
\newtheorem{Cor}{Corollary}
\newtheorem{Def}{Definition}
\newtheorem{Exam}{Example}
\newtheorem{Alg}{Algorithm}
\newtheorem{Prob}{Problem}
\newtheorem{Rem}{Remark}

\author{\IEEEauthorblockN{Shunqing Zhang$^{\dagger}$,~\IEEEmembership{Senior Member,~IEEE}\\
Qingqing Wu$^{\ddagger}$,~\IEEEmembership{Student Member,~IEEE}, \\
Shugong Xu$^{\dagger}$,~\IEEEmembership{Fellow,~IEEE}, \\
and Geoffrey Ye Li$^{\ddagger}$,~\IEEEmembership{Fellow,~IEEE},} 
\IEEEauthorblockA{$^{\dagger}$Intel Labs, Beijing, 100080, China\\
$^{\ddagger}$Georgia Institute of Technology, Atlanta, GA, 30332, USA\\
Emails: \{shunqing.zhang, shugong.xu\}@intel.com, \{qingqing.wu, liye\}@ece.gatech.edu}
\thanks{}}
\maketitle

\begin{abstract}
With years of tremendous traffic and energy consumption growth, green radio has been valued not only for theoretical research interests but also for the operational expenditure reduction and the sustainable development of wireless communications. Fundamental green tradeoffs, served as an important framework for analysis, include four basic relationships: spectrum efficiency (SE) versus energy efficiency (EE), deployment efficiency (DE) versus energy efficiency (EE), delay (DL) versus power (PW), and bandwidth (BW) versus power (PW). In this paper, we first provide a comprehensive overview on the extensive on-going research efforts and categorize them based on the fundamental green tradeoffs. We will then focus on research progresses of 4G and 5G communications, such as orthogonal frequency division multiplexing (OFDM) and non-orthogonal aggregation (NOA), multiple input multiple output (MIMO), and heterogeneous networks (HetNets). We will also discuss potential challenges and impacts of fundamental green tradeoffs, to shed some light on the energy efficient research and design for future wireless networks.
\end{abstract}



\newpage

\section{Introduction}\label{sec:intro}
With years of tremendous traffic and energy consumption growth, green radio \cite{GT:09} has been valued not only for theoretical research interests but also for the operational expenditure reduction and the sustainable development of wireless communications. \textcolor{black}{Before attracting extensive research interests, several pioneer works \cite{CF:xia2004latency,CF:cui2005energy,JR:heo2005energy,
CF:bae2007tradeoff,JR:neely2007optimal,CF:leow2007delay,CF:brand2008delay,
JR:meshkati2009energy,JR:ban2009capacity,JR:chen2009hybrid,CF:richter2009energy,
CF:predojev2010energy,CF:gorce2010energy,CF:ra2010energy} have been devoted to investigate the tradeoffs between energy, bandwidth, deployment, and delay independently for different scenarios.  }
 \textcolor{black}{In order to provide a holistic view on the above issues, fundamental green tradeoffs have been summarized in \cite{JR:Chen11}.
  Within this framework, four basic tradeoff relationships among key performance metrics and parameters have been analyzed, including spectrum efficiency (SE) versus energy efficiency (EE), deployment efficiency (DE) versus energy efficiency (EE), delay (DL) versus power (PW), and bandwidth (BW) versus power (PW) relations.} Great amount of efforts have been made to fully utilized the fundamental tradeoffs through international research organizations or collaborative projects. For example, GreenTouch \cite{GT:09}, a worldwide research consortium aiming to improve the EE of communication systems, has established a dedicated project ``Green Transmission Technologies (GTT)'' to investigate the underlying tradeoffs for practical networks. Energy Aware Radio and network TecHnologies (EARTH) \cite{EARTH:D2.3} project has applied these tradeoffs to demonstrate the final deliverables, including energy efficient network architecture and deployment strategies.
\textcolor{black}{Besides these projects,  IEEE Communications Society have also recognized several books \cite{BK:hossain2012green,BK:wu2012green,BK:yu2012green,BK:obaidat2012handbook} and Journal/Magazine Special Issues featured on green communications as best readings \cite{br_green}, which aim at providing readers a comprehensive coverage on green communications from the state-of-the-art theoretical research results to industrial applications.}

Along with the dramatic traffic explosion, it is also promising to see that the next generation (5G) wireless communications shall include people-to-machine and machine-to-machine communications \cite{METIS:12} in order to facilitate more flexible networked social information sharing. As a result, numerous sensors, accessories, or even tools may become the communication entities and the associated running applications over wireless networks will diverge. As reported in \cite{5g:14}, wireless communications shall support up to millions of applications and billions of subscribers by year 2020, which is nearly 100 times of today's network. Consequently, with the surprisingly expanding demands for wireless transmission and supporting equipment,  the network power consumption is no longer sustainable and the green radio technology becomes essential \cite{JR:I14}. A flagship 5G research project from European Union, named Mobile and wireless communications Enables for the Twenty-twenty Information Society (METIS), has introduced the green metric, ECON-B/AU, as a basic system performance measure \cite{METIS:12} and claimed to have 1,000 times EE improvement between 2010 and 2020. Meanwhile, a green and soft future for 5G systems from the operator's viewpoint has been presented in  \cite{JR:I14} and vendors are managing to prolong the battery life of devices in 5G systems by a factor of ten \cite{5g:14}.

Due to tremendous research efforts and lots of progresses in green radio in the past few years, a number of surveys and tutorials have been contributed to EE improvement for cellular networks. In particular, a comprehensive study on the energy efficient schemes for orthogonal frequency division multiple access (OFDMA), multiple input multiple output (MIMO), and relay networks has been provided in \cite{JR:Li11}. In \cite{JR:Bianzino11},  a complete overview for both wired and wireless networks has been conducted. A dedicated discussion on the energy efficient wireless communications has been characterized in \cite{JR:Feng13}. More recently, the literature surveys for EE improvement have nailed down to several specific technical areas in cellular networks, including radio resource management \cite{JR:Rao14}, dynamic resource provisioning \cite{JR:Wu15}, base station sleeping \cite{JR:Budzisz14} as well as low power hardware techniques \cite{JR:Davaslioglu14}. Nevertheless, the surveys above have focused on specific energy efficient schemes. \textcolor{black}{In this article, we will focus on: 1) theoretical progresses on fundamental green tradeoffs, 2) joint analysis with energy efficient solutions and hardware characteristics, and 3) design impacts for future green networks.  Fundamental green tradeoffs as well as the joint analysis and design impacts have also been identified in \cite{JR:Hasan11}. However,
\cite{JR:Hasan11} emphasizes cognitive radio and cooperative techniques while we focus on the main 4G technologies and the promising 5G technologies, which are fundamentally different.}

Since we are at the transition stage from 4G to 5G networks nowadays, it is critical for us to understand the on-going energy efficient solutions for 4G networks and identify some forward-looking research challenges on green 5G network design. Hence, after we summarize the extensive on-going research efforts and categorize them based on  fundamental green tradeoffs in this paper, we will focus on research progresses in 4G networks, such as orthogonal frequency division multiplexing (OFDM), MIMO, and heterogeneous networks (HetNets). We then discuss potential impacts and identify research challenges of the fundamental green tradeoffs in 5G networks, where the most promising features, including non-orthogonal aggregation (NOA) \cite{CF:Bharadia13, JR:Feng131, JR:Ding14}, massive MIMO (M-MIMO) \cite{JR:Lu14} and an ultra dense network (UDN) \cite{JR:Yunas15}, are re-evaluated under the same framework, respectively.

The rest of the paper is organized as follows. Section \ref{sec:cur_status} will provide a comprehensive overview on the current research progresses of the fundamental green tradeoffs. Based on that, Sections~\ref{sec:ofdm}, \ref{sec:mimo},  and \ref{sec:hetnet} will discuss three types of technologies, including OFDM and NOA, regular and massive MIMO, HetNets and UDNs, respectively. Other energy efficient technologies are also summarized in Section~\ref{sec:mmWave}. We present a heterogeneous design framework for energy efficient networks in Section \ref{sec:hetero} and conclude the paper in Section \ref{sec:con}.

\section{Fundamental Green Tradeoffs}\label{sec:cur_status}
The fundamental green tradeoffs in an additive white Gaussian noise (AWGN) channel have been proposed in \cite{JR:Chen11}. Following the conventional notations, we denote $P, W, N_0, \tau, R, C$ as the total transmit power, the total bandwidth, the additive noise power density, the transmission delay\textcolor{black}{\footnote{\textcolor{black}{Since we mainly focus on the physical layer transmission, we only adopt the transmission delay here to characterize the delay-power tradeoff which can be regarded as a theoretical limit. }}}, the rate requirement, and the deployment cost. The four fundamental tradeoff relations in an AWGN channel can be expressed as \cite{JR:Chen11},
\begin{eqnarray}
& \textrm{SE-EE Tradeoff:} & \ \eta_{EE} (\eta_{SE})= \frac{\eta_{SE}}{\left(2^{\eta_{SE}} - 1 \right) N_0} \label{eqn:SE-EE}\\
& \textrm{DE-EE Tradeoff:} & \ \eta_{EE} (\eta_{DE})= \frac{\eta_{DE} \cdot C}{\left(2^{\frac{\eta_{DE} \cdot C}{W}} -1 \right) W N_0} \label{eqn:DE-EE}\\
& \textrm{BW-PW Tradeoff:} & \ P(W) = W N_0 \left(2^{\frac{R}{W}} - 1 \right) \\
&  \textrm{DL-PW Tradeoff:}  & \ P(\tau) = W N_0 \left(2^{\frac{1}{W \tau }} - 1 \right)
\end{eqnarray}
where
\begin{eqnarray}
\eta_{SE} & = & \log_2 \left(1 + \frac{P}{W N_0}\right) \\
\eta_{EE} & = & \frac{W}{P} \log_2 \left(1 + \frac{P}{W N_0}\right) \\
\eta_{DE} & = & \frac{W}{C} \log_2 \left(1 + \frac{P}{W N_0}\right)
\end{eqnarray}
denote SE, EE and DE, respectively\textcolor{black}{\footnote{\textcolor{black}{We note that area energy efficiency is also another important performance metric for cellular networks. Since it is more related with the cell size, we will emphasize it in heterogeneous networks in Section V. }}.}

\begin{figure}
\centering
\includegraphics[width = 6 in]{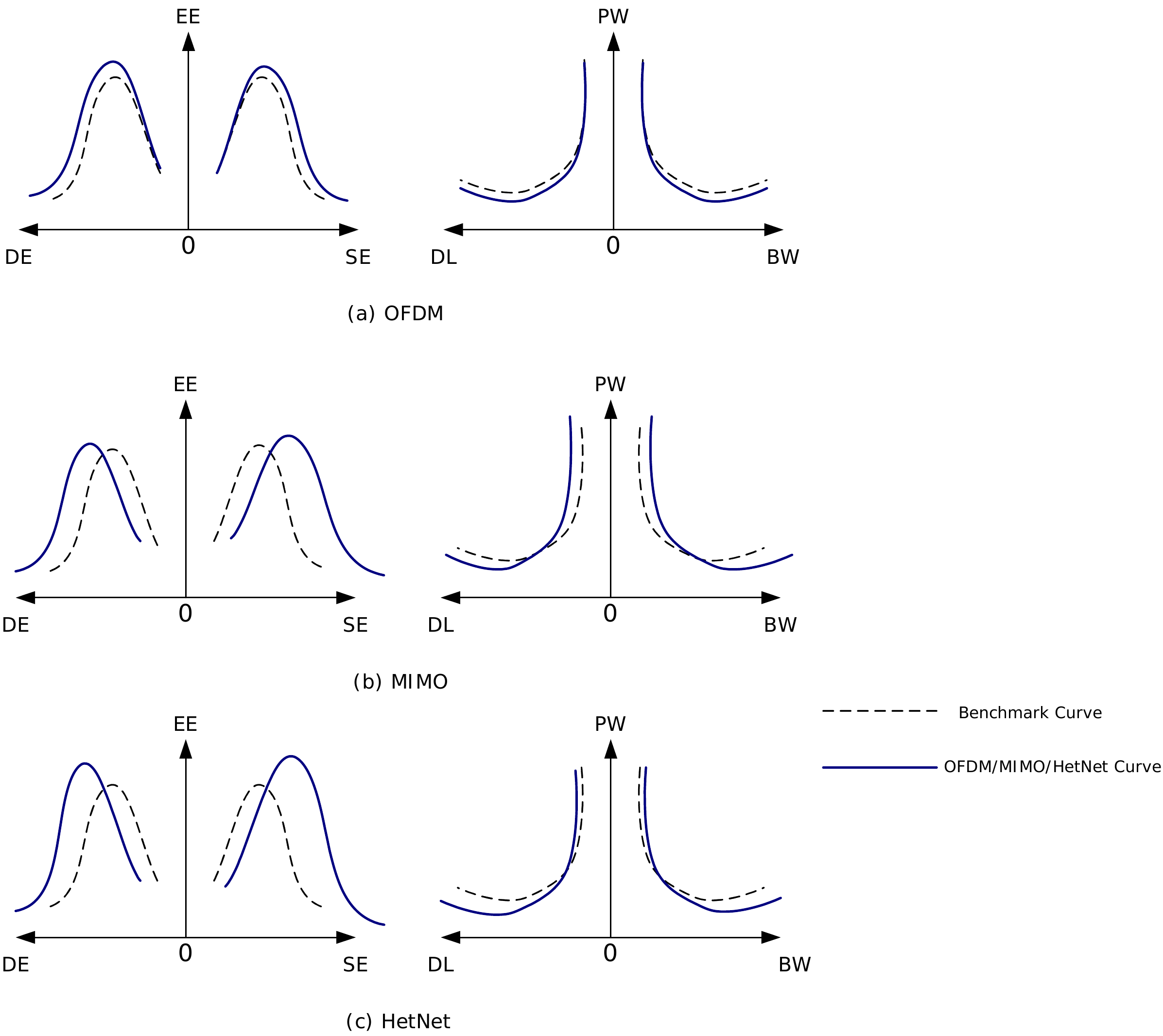}
\caption{\textcolor{black}{Fundamental green tradeoffs for benchmark systems and their evolutions to OFDM, MIMO, and HetNets with circuit power consideration. The dashed curves show fundamental green tradeoffs for benchmark systems as illustrated in \cite{JR:Chen11}. The curves in a), b), and c) stand for fundamental green tradeoffs in OFDM, MIMO, and HetNets, and the corresponding analysis and demonstration can be found in Section~\ref{sub_sec:ofdm}, \ref{sub_sec:mimo},  and \ref{sub_sec:hetero}, respectively.} }
\label{fig:fun_tra}
\end{figure}

In the practical systems, however, the tradeoff curves may behave differently. For example, if the circuit power consumption, $P_{c}$, is considered in the EE evaluation, then $\eta_{EE} = \frac{W}{P + P_{c}} \log_2 \left(1 + \frac{P}{W N_0}\right)$ and the corresponding SE-EE/DE-EE tradeoff curves will be with a bell shape \cite{JR:Chen11} as shown in Fig. \ref{fig:fun_tra}. BW-PW/DL-PW tradeoffs under the circuit power assumptions have been also discussed in \cite{JR:Chen11}.  In particular, several open issues have been raised in \cite{JR:Chen11}, including tradeoff analysis for multi-cell systems and HetNet architectures.

\begin{figure}
\centering
\includegraphics[width = 6 in]{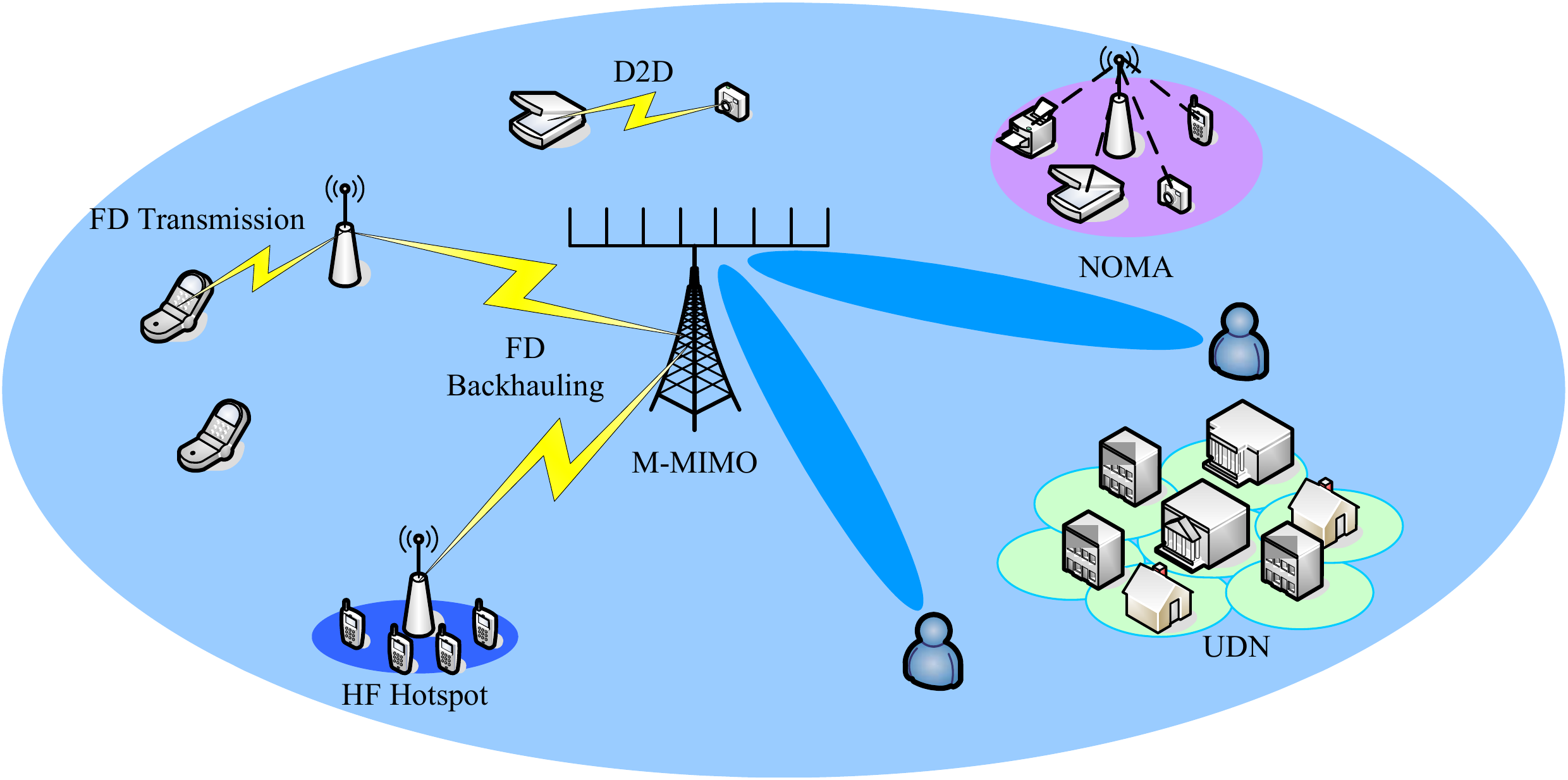}
\caption{A summary of key technologies for 5G communication systems and their application scenarios. For example, NOA (such as NOMA, D2D and FD), as a novel breakthrough for physical layer, is specialized for massive low-rate terminals. M-MIMO is designed to improve the throughput via smart designed narrow beams and a UDN technology is targeted for hotspots with dense users. }
\label{fig:5g_scena}
\end{figure}

Four fundamental green tradeoffs can actually stretch together key network performance/cost indicators and connect the technologies toward green evolution in different aspects, including network planning, resource management, and physical layer transmission scheme design. Moreover, guided by the fundamental green tradeoffs, years of tremendous efforts have been put into the green technologies for the existing as well as future cellular networks, especially for 4G and 5G communications. As we have indicated before, OFDM, MIMO, and HetNets are often regarded as the most important features for 4G networks, and have attracted significant attention for green communications in the past several years. Meanwhile, to provide 1,000 times throughput improvement, 100 times delay reduction and 100 times connections for 5G networks while keeping the energy consumption sustainable, energy efficient communication covers all promising 5G technologies, including NOA, M-MIMO, and UDN as shown in Fig. \ref{fig:5g_scena}, and has become an important research topic recently \cite{JR:I14}. The fundamental tradeoff curves have been characterized, some open issues in \cite{JR:Chen11} have been resolved, some physical limitations have been identified. In the following sections, we will discuss the research achievements of the above features through the following three aspects. Firstly, we summarize the analytical results of fundamental green tradeoff curves for different features. Then, guided by the fundamental green tradeoffs, we analyze the existing energy efficient solutions in detail and investigate the power minimization problems when affecting the BW-PW/DL-PW tradeoffs. Finally, we raise some future research issues related to the fundamental green tradeoffs.


\section{OFDM and NOA} \label{sec:ofdm}
Time-frequency resources utilization is one of the most important technologies in communication networks, where the orthogonal utilization using OFDM is applied in 4G systems and the non-orthogonal extension using NOA has been proposed for 5G systems. In this section, we summarize the energy efficient solutions for OFDM and NOA technologies based on the fundamental green tradeoffs.

\subsection{OFDM} \label{sub_sec:ofdm}
OFDM, as a landmark technology for 4G systems, is originally proposed to deal with the frequency selectivity of wireless channels due to the multi-path fading in broadband transmission. Fundamental green tradeoffs for OFDM actually manage to figure out the inner relationship between the achievable system performance and the consumed power resource, and the main results are as shown below.

\subsubsection{Theoretical Achievements}
Theoretical analysis on fundamental green tradeoffs for OFDM systems is rather straight forward. The ergodic capacity for OFDM can be expressed as
\begin{eqnarray}
C_{OFDM} = W \cdot \mathbb{E}_{\{h_j\}} \left[ \sum_{j=1}^{N_s}\log_2 \left(1 + \frac{P}{W N_0} \|h_j\|^2\right) \right], \label{eqn:ofdm}
\end{eqnarray}
where $j$ is the subcarrier index, $N_s$ is the total number of subcarriers, and $h_j$ denotes the channel frequency response at the $j^{th}$ subcarrier. From $\eta_{SE} = C_{OFDM}/W$, $\eta_{EE}(\eta_{SE}) = \eta_{SE} W / P$ and $\eta_{DE}(\eta_{SE}) = \eta_{SE} W / C$, the EE/DE of OFDM systems increases while SE decreases with the bandwidth, which means OFDM provides better EE/DE in the wide band area at the expense of losing SE. Meanwhile, since the ergodic capacity, $C_{OFDM}$, is an increasing function with respect to the bandwidth, $W$, it has better BW-PW and DL-PW tradeoffs, especially in the low SE regime.

In the practical scenario, due to complicated signal processing and high peak-to-average power ratio (PAPR) in OFDM transmission \cite{BK:Nee}, the circuit power consumption, $P_{c}$, is never small. However, with the development of hardware processing capability, especially with the help of fast Fourier transform (FFT), traditional power-hungry OFDM technology can be implemented with reasonable hardware cost and manageable power consumption \cite{JR:Wang00}. \textcolor{black}{As a result, the fundamental green tradeoff curves for practical OFDM systems can be shown in Fig. \ref{fig:fun_tra}(a), which provides a better EE/PW performance if compared with the baselines.}

The above theoretical framework has been extended to OFDMA scenarios \cite{derrick_harvest13,ng2012energy3,ng2012energy1,JR:ng2012energy2}. In OFDMA systems, sub-channelization of OFDM technology allows base stations to opportunistically select the most suitable users for each sub-channel and timely adjust the coding and modulation schemes to improve EE performance \cite{JR:Xiong11, JR:Xiong12, JR:Luo14, JR:qing1,JR:qing15tx_rx}. For example, downlink energy efficient power allocation schemes have been proposed in \cite{JR:Xiong11}, where it has been shown that the maximum achievable EE is strictly quasi-concave in terms of SE for a sufficient large number of subcarriers. The analytical framework has been extended to uplink OFDMA systems in \cite{JR:Xiong12}, where the minimum individual EE maximization guarantees the worst user performance. In addition to the energy consumption at the transmitters, that for the receivers has been ignored before and is also considered in \cite{JR:Luo14,JR:qing1,JR:qing15tx_rx} when formulating the systematic EE.

\subsubsection{Energy Efficient Solutions}
Significant research efforts have been devoted to obtain energy efficient design for OFDM systems. In the downlink scenario,  the initial work on the point-to-point energy efficient link adaptation scheme over frequency-selective channel can be traced back to \cite{JR:Miao10}, where the throughput-per-Joule EE metric considering circuit power consumption has been first proposed and the iteratively subcarrier and power allocation has been analyzed. The result has been extended to EE-oriented link adaptation for multiuser scenarios in \cite{JR:Xu12}, where the practical pilot power consumption and the induced channel estimation errors have been also studied. It has been concluded that utilizing the maximum power for pilots and data transmission is not always optimal for EE-oriented design. In \cite{JR:Ren14}, the multiuser fairness issues for energy efficient design have been further investigated, where proportional rate constraints have been applied in the EE formulation. It has been proved in \cite{JR:Ren14} that the EE expression is strictly quasi-concave in the proportional rate parameter, $\lambda$,  defined as the rate requirement over the proportional rate constraint. Thus, the EE-optimal algorithms in  \cite{JR:Miao10} and \cite{JR:Xu12} can be applied here. In general, the above works have shown that the energy efficient downlink transmission strategies can improve the EE performance significantly with a relatively small SE loss. For OFDM networks with delay-sensitive traffic, the concept of {\em effective capacity} \cite{JR:Tang08} can be used to formulate the EE or power optimization problem if perfect channel state information (CSI) is known at the transmitter. In \cite{JR:Xiong13},  delay-guaranteed constraints have been transformed into the transmission rate requirements. Therefore, the EE-oriented design methodology as proposed in \cite{JR:Xu12} has been directly adopted to obtain EE-optimal strategies. A more complete DL-PW relation for OFDM systems has been analyzed in \cite{JR:Zhang13} and \cite{JR:Ji14}, where the effective capacity to link the delay performance and the power requirements has been used. The heterogeneous queueing delay and power tradeoff coupled with the imperfect CSI has been discussed in \cite{JR:Lau12}, where the Lyapunov stochastic stability analysis \cite{JR:Cui12_IT}  has been applied to develop the corresponding dynamic back-pressure power control algorithm to achieve the optimal DL-PW relation.

\begin{table}
\tabcolsep 2mm
\renewcommand{\arraystretch}{2}
\footnotesize
\caption{Taxonomy of Energy Efficient OFDM Schemes based on the Fundamental Green Tradeoffs} \label{table:ofdm} \centering
\begin{tabular}{c|c|c|c|c}
\hline
\hline
Ref. & Main Tradeoff & Technical Area & DL/UL & Contribution \\
\hline
\hline
\cite{JR:Xiong11, JR:Luo14} & SE-EE & Theoretical Analysis & DL & Characterize the tradeoff curves with circuit power\\
\hline
\cite{JR:Xiong12} & SE-EE & Theoretical Analysis & DL/UL & Characterize the tradeoff curves with circuit power\\
\hline
 \cite{JR:Miao10, JR:Xu12} & SE-EE, BW-PW & Resource Management & DL & Energy efficient link adaptation with circuit power\\
\hline
\cite{JR:Ren14} & SE-EE & Resource Management & DL & Energy efficient multiuser link adaptation with user fairness\\
\hline
\cite{JR:Xiong13} & DL-PW, SE-EE & User Scheduling & DL & Energy efficient multiuser scheduling with delay constraints\\
\hline
\cite{JR:Zhang13, JR:Ji14} & DL-PW & Theoretical Analysis & DL & Characterize the tradeoff curves with circuit power\\
\hline
\cite{JR:Lau12} & DL-PW & User Scheduling & DL & Multiuser scheduling under heterogenous queueing delay\\
\hline
\cite{JR:Miao12} & SE-EE & Resource Management & UL & Distributed power control for multiuser EE maximization \\
\hline
\cite{JR:Dechene14} & DL-PW & Resource Management & UL & Distributed power control with ARQ delay constraints\\
\hline
\cite{JR:Buzzi12} & SE-EE & Resource Management & UL &  Distributed power control for multiuser multi-cell systems \\
\hline
\cite{JR:Jiang13, JR:Joung14} & DE-EE & Implementation & DL/UL & Characterize the impacts of hardware components on EE\\
\hline
\hline
\end{tabular}
\end{table}
In the uplink, naturally we need to deal with the EE or power related issues in multiuser scenarios through efficient user scheduling and power control \cite{JR:qing15tx_rx}. However, to obtain the optimal solutions, complicated iterative search and global information change for each time interval is required, which is with prohibited complexity and signaling overhead. To deal with these issues, a distributed power control algorithm has been proposed in \cite{JR:Miao12}, which relies on the localized channel states of all subchannels and the history of data transmission and power consumption. It has also been proved in \cite{JR:Miao12} that the existing water-filling power allocation approaches can still be applied for a given transmit power and selected subcarrier sets. A more practical design for 4G uplink systems with synchronous HARQ constraints can be found in \cite{JR:Dechene14}, where the block interval based margin adaptation scheme is applied to guarantee the ARQ delay and minimize the potential transmit power. However, the above approaches require the frequently updated network information to obtain the optimal/sub-optimal strategies and may not be applicable in the multi-cell environments due to the limited backhaul capacity for inter-cell communications.  To address this issue, potential games have been formulated in \cite{JR:Buzzi12} for energy efficient power control and subcarrier allocation in uplink multi-cell OFDMA systems. With the noncooperative game formulation, each uplink user plays an signal-to-interference-and-noise ratio (SINR) maximization game to determine the subcarrier allocations and a distributed power control game to maximize the EE based on their local network information. The game solution in \cite{JR:Buzzi12} has been proved to converge and achieve order-wise EE improvement compared with benchmark schemes. Table \ref{table:ofdm} provides a summary of the above literatures. \textcolor{black}{In summary, \cite{JR:Xiong11, JR:Xiong12,JR:Luo14,JR:Miao10,JR:Ren14,JR:Miao12,JR:Buzzi12} are more applicable to large scale systems that pursue high EE and SE while other figures of merit, such as delay and bandwidth, are ignored. In contrast, the approaches in \cite{JR:Lau12,JR:Zhang13, JR:Ji14,JR:Dechene14,JR:Xiong13,JR:Xu12} are preferred by delay sensitive applications and power limited wireless networks. Some details and guidance on implementing energy efficient solutions to realistic systems can be found in \cite{JR:Jiang13, JR:Joung14}. }

\subsubsection{Future Research Issues}
OFDM exploits EE potentials in the time-frequency domain and the corresponding impacts on hardware components (e.g. power amplifiers) are not difficult to characterize in general \cite{JR:Jiang13, JR:Joung14}. However, the current EE metric design in the OFDM configuration is not applicable in evaluating multiuser scenarios. Usually, users with a lower circuit power are more energy efficient than those with a higher circuit power, which generates a fairness issue. Another challenge is inter-cell interference modeling. Due to the distributed user scheduling, the splitted power as well as the modulation and coding schemes for each sub-channel/subcarrier are generally unpredictable, and the induced interference will be uncontrollable at the user side. Meanwhile, as illustrated in \cite{BK:Nee}, the radiated power and signal quality of OFDM systems depend on the instantaneous PAPR, which eventually makes generated interference (to neighboring cells) highly dynamic. Hence, further investigation on the following research topics is still expected.
\begin{itemize}
\item{\em How to design energy efficient schemes with heterogeneous circuit power consumptions?} Due to various types of terminals, the circuit power consumptions may differ dramatically, which will affect the EE metric evaluation and the fairness among users \cite{JR:Feng13}. Therefore, new formulation of the EE design problem with fairness consideration for near-far effect and heterogeneous circuit power consumption will be essential.
\item{\em How to design energy efficient schemes with highly dynamic inter-cell interference?} One possible solution to solve this issue is to control the potential inter-cell interference sources, e.g. through low PAPR signal design or low noise spectrum filtering. Another possible way is to facilitate inter-cell interference information exchange \cite{JR:Rao14}, e.g. to allow multi-cell cooperation or coordinated interference cancellation. For either approach, additional research efforts on energy efficient schemes will be worthwhile.
\end{itemize}

\subsection{Non-Orthogonal Aggregation}
The extension of OFDM in the time-frequency domain is not straight forward since the existing 4G systems have fully utilized the resources in the orthogonal sense. NOA intends to provide time-frequency domain aggregation with the non-orthogonal technology, which has been recently proposed as a key enabling technology for 5G.

\subsubsection{Theoretical Achievements}
\textcolor{black}{Typical NOA technologies include full duplexing (FD) \cite{CF:Bharadia13}, device-to-device (D2D) \cite{JR:Feng131}, and Non-Orthogonal Multiple Access (NOMA) \cite{JR:Ding14}, where FD focuses on the self-interference cancellation and the D2D and NOMA deal with inter-user interference.} Instead of analyzing a single transmitter-receiver pair as in the previous cases, the ergodic capacity of NOA technology can be described by a sum capacity of all transmission pairs and can be expressed as,
\begin{eqnarray}
C_{NOA} = W \cdot \mathbb{E}_{\{h_k, h_{k}^{n}\}} \left[\sum_{k=1}^{N_p} \log_2 \left( 1 + \frac{P_k \|h_k\|^2}{W N_0 + \xi_k \sum_{n=1, n \neq k}^{N_p}  P_{n} \|h_{k}^{n} \|^2}\right) \right],
\end{eqnarray}
where $k$ is the index of transmission pair, $N_p$ is the total number of transmission pairs and $P_{k}$ is the equivalent transmit power of the $k^{th}$ transmission pair. $h_k$ and $h_{k}^{n}$ denote the channel gains of the $k^{th}$ pair and  the interference link to the $k^{th}$ pair generated from the $n^{th}$ transmission pair, respectively. $\{\xi_k\}$ is the equivalent interference cancellation capability of the $k^{th}$ pair, which can be regarded as a joint effect of the NOA transmission scheme, the interference cancellation technique of the $k^{th}$ receiver \cite{JR:Hong142}, and other power management techniques \cite{JR:Kim12}. Consequently, the sum SE of NOA, $\eta_{SSE}$, can be calculated via $\eta_{SSE} = C_{NOA} / W$ and the maximum achievable sum SE, $\eta_{SSE}^{\star}$, is given by
\begin{eqnarray}
\eta_{SSE}^{\star} = \max_{P_k \geq 0, \  \sum P_k = P} \mathbb{E}_{\{h_k, h_k^{k'}\}} \left[ \sum_{k = 1}^{N_p} \log_2 \left(1 + \frac{P_k \|h_k\|^2}{W N_0 + \xi_k \sum_{n =1 \in, n \neq k}  P_{n} \|h_k^{n}\|^2}\right) \right].
\end{eqnarray}
With proper power arrangement and advanced inference cancellation schemes, NOA is able to achieve a better sum SE compared with the orthogonal transmission schemes. The sum EE, $\eta_{SEE}$, can be obtained similarly and  $\eta_{SEE}$ grows proportionally with $\eta_{SSE}$ for fixed total transmit power $P$, which results in a better SE-EE tradeoff curve \cite{JR:Nguyen14, Saito:13, JR:Zhou14}.

The additional deployment cost for the NOA scheme is generally controllable since no additional site renting or infrastructure upgrading is required. As a result, a better DE-EE tradeoff relation will not be surprising. In general, since NOA provides better utilization of time-frequency resources and reduces the potential queueing delay by creating more virtual transmission pairs, better power related tradeoffs  can be obtained compared with the orthogonal solutions. In the practical networks, the number of required processing chains scales with the number of the active transmission pairs in the NOA scheme and a linear growth of static circuit power will be possible. In that case, we can no longer achieve a better sum EE as in the ideal scenario and the corresponding BW-PW and DL-PW tradeoffs will be affected.

\subsubsection{Energy Efficient Solutions} \label{sec:ee_noa}
EE enhanced solutions for NOA have been widely discussed recently and one of the promising directions is to control the PAPR and out of band leakage to limit non-orthogonal interference. Several advanced waveform techniques have been proposed to support the energy efficient NOA transmission. A flexible modulation scheme, called generalized frequency division multiplexing (GFDM), has been proposed in \cite{JR:Michailow14} to control the out of band emission, which makes fragmented spectrum and dynamic spectrum allocation feasible without severe interference. Filter-bank multi-carrier (FBMC) is with a low PAPR and fits for broadband transmission. Furthermore, it can be extended to multi-stream scenario as shown in \cite{JR:Caus14}, which makes it possible for NOA applications. Other types of tunable OFDM, such as DFT-precoded OFDM, a punctured variable-length prefix OFDM and universal filtered multi-carrier (UFMC) as summarized and compared in \cite{JR:Andrews14} and \cite{JR:Wunder14}, provide various choices for energy efficient NOA waveform design.

Another promising direction is to develop reconfigurable hardware platform for the NOA processing. Nowadays, effective interference rejection schemes contain a mixture of spatial, analog, and digital domain cancellation in order to compensate non-linear effects and other special effects such as oscillator noise \cite{CF:Bharadia13}. However, this approach typically requires fine-grained tuning of some engineering parameters and the corresponding interference cancellation is often complexity and power hungry. If different NOA schemes need to be supported, a brute-force implementation to configure multiple hardware links definitely consumes prohibitive power and hardware cost. Therefore, an energy efficient NOA implementation using reconfigurable platform is promising \cite{JR:Dai15} and the associated unified frame structure design \cite{JR:Levanen14} and low-overhead protocol \cite{JR:Feng14} to reduce the switching latency of multiple NOA schemes are also important.

\textcolor{black}{D2D communications have been envisioned as a key technology for 5G networks \cite{JR:liu2014device,JR:Feng14,JR:mach2015band,JR:wei2014enable,JR:feng2015mode,CF:akkarajitsakul2012mode,CF:della2015mode}. In particular, proximity users are allowed to communicate directly without going through the base station \cite{JR:wei2014enable}. The SE-EE tradeoff for D2D communications underlaying cellular networks has been investigated in \cite{JR:Zhou14}. It has been pointed out  in \cite{JR:Zhou14} that increasing transmission power beyond the power for maximum EE brings little SE improvement but signiﬁcant EE loss, especially when the interferences between D2D users are not so strong.   The overall system EE and the individual user EE are both optimized in \cite{CF:hoang2015energy} via joint channel assignment and power allocation under the assumption that there is no interference among D2D pairs. Then, this assumption is relaxed in \cite{CF:hoang2015dual} where it has been shown that the maximum EE is enhanced significantly but sensitive to the number of D2D pairs due to frequency reuse. Beside resource allocation, mode switching is also attractive for realizing energy efficient D2D communications \cite{JR:feng2015mode,CF:akkarajitsakul2012mode,CF:della2015mode}. Specifically, D2D users are allowed to switch between cellular mode, reuse mode, and dedicated mode, to optimize the system performance while without causing severe interference to cellular users.
}

\textcolor{black}{Another appealing technology is full-duplex \cite{JR:wei2015full,JR:nguyen2013precoding,JR:liu2015energy,JR:chen2015spectral,JR:maso2015energy,JR:zeng2015full,CF:dongenergy2015,
JR:alenergy2016,JR:jia2015spectrum,CF:el2015energy,sun2015multi,ng2015power}.
A comprehensive EE analysis of the FD and half-duplexing (HD) amplify and-forward (AF) has been provided in \cite{JR:wei2015full}, where it has been shown that FD
with passive suppression is more energy efficient than HD. In addition, when the transmission power is high,  the system EE can be further improved by exploiting analog cancellation.  EE and SE tradeoff as well as resource allocation has been investigated in FD multiuser MIMO \cite{JR:nguyen2013precoding}, one-way relay \cite{JR:liu2015energy}, and  two-way relay \cite{JR:chen2015spectral} systems. Then, energy recycling is also integrated into FD systems to achieve high EE and SE \cite{JR:maso2015energy,JR:zeng2015full,CF:dongenergy2015}, where a part of energy that is used for information transmission by a transmitter can be harvested and reused. Beside the physical layer aspect, novel solutions and protocol enhancements are also needed at higher layers to achieve the true benefits of FD technology.  An energy-efficient medium
access control (MAC) protocol has been proposed in \cite{JR:alenergy2016} to reduce the transmission power of data and acknowledgement (ACK) packets.  Significant EE gain has been demonstrated in \cite{JR:alenergy2016} and but higher layer gains of FD technology heavily rely on self-interference cancellation mechanisms at the physical layer.
}

\subsubsection{Future Research Issues} 
\begin{figure}
\centering
\includegraphics[width = 6 in]{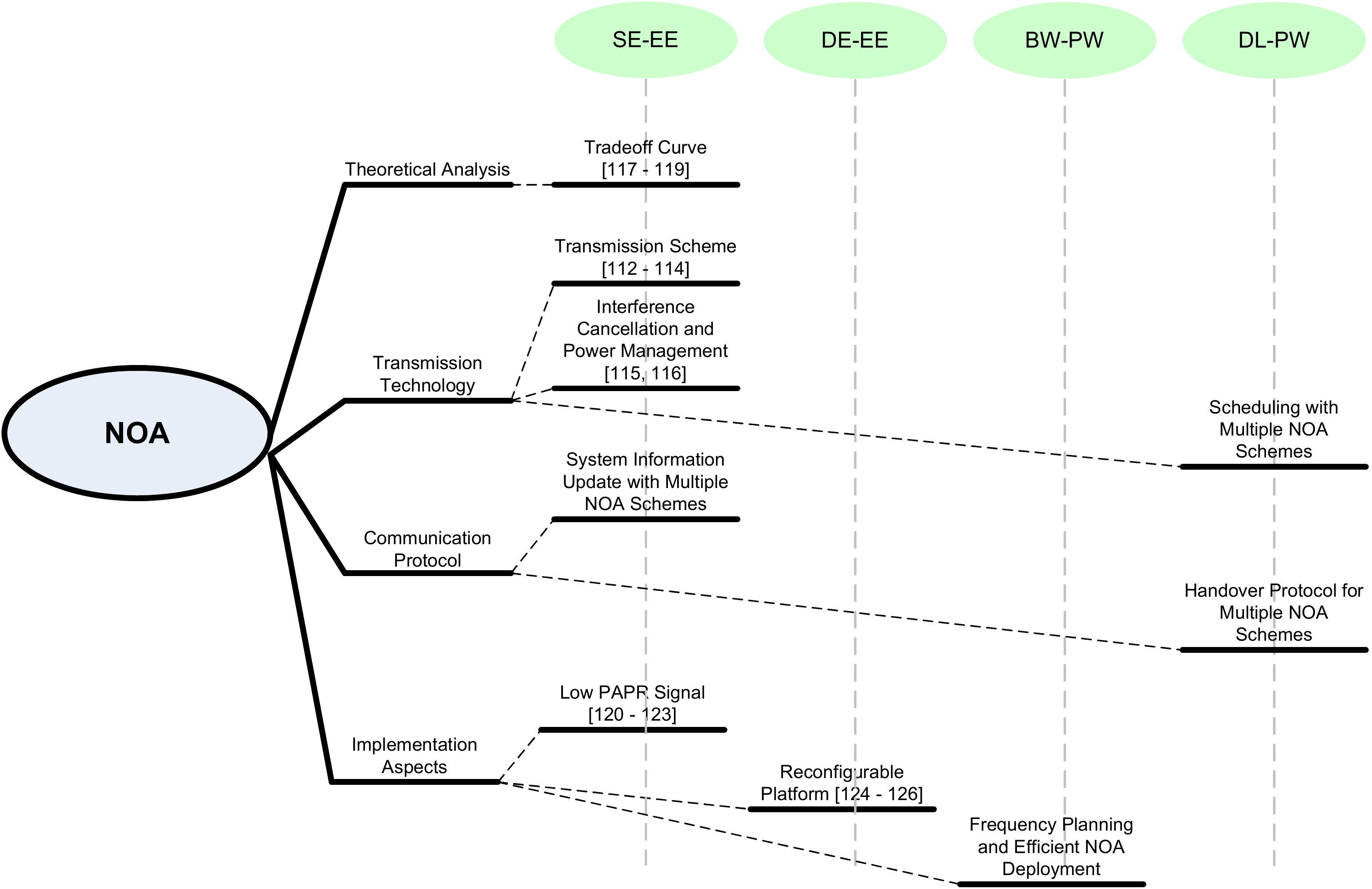}
\caption{A research roadmap for energy efficient NOA technology based on the fundamental green tradeoffs.}
\label{fig:noa}
\end{figure}

Current energy efficient NOA transmission focuses on the implementation aspect, where the hardware friendly signal and reconfigurable techniques have been discussed in the previous part.  Inspired by the energy efficient schemes for orthogonal access systems, we may consider the questions for energy efficient NOA as concluded in Fig. \ref{fig:noa}.

\textcolor{black}{\begin{itemize}
\item{\textcolor{black}{\em How to deploy NOA technologies for different users more energy efficiently?}} \textcolor{black}{The deployment problem is also critical for energy efficient NOA transmission. As an example, NOMA is more suitable for users with significant pathloss difference in order to improve the performance of successive interference cancellation \cite{JR:Ding14} while FD prefers closer transmit-receive distance with lower transmit power due to limited self-interference cancellation capability \cite{JR:wei2015full,JR:nguyen2013precoding,JR:liu2015energy}. Hence, to deploy NOA technologies according to the serving scenarios can greatly improve the overall EE performance and requires further investigation.}
\item{\textcolor{black}{ \em How to design energy efficient scheduling strategies for users with multiple NOA schemes?}} \textcolor{black}{After deploying different NOA schemes for different users, how to allocate resources among them becomes a critical problem \cite{JR:Andrews14,JR:Wunder14,JR:Dai15}. As we have explained in Section~\ref{sec:ee_noa}, it is power and hardware costly to support multiple NOA schemes using multiple hardware links. An energy efficient strategy is to schedule users with the same NOA scheme together. However, the associated problems to guarantee the delay requirements and avoid frequent reconfiguration of hardware are more difficult and still open.}
\end{itemize}}

\section{MIMO} \label{sec:mimo}
In the spatial domain, regular MIMO has been applied in 4G systems, and in order to support order-wise throughput enhancement for 5G systems, a massive extension of MIMO technology has been proposed. In this section,  we discuss fundamental green tradeoffs for regular and massive MIMO systems, and identify several critical open issues from the EE perspectives.

\subsection{Regular MIMO} \label{sub_sec:mimo}
MIMO aims to improve the SE by using multiple transmit and receive antennas. Meanwhile, it also incurs higher hardware cost or more operating power consumption. Fundamental green tradeoff study provides a more comprehensive view for MIMO technology especially for energy or power related metrics.

\subsubsection{Theoretical Achievements}
The detailed SE-EE tradeoff has been analyzed in \cite{JR:Heliot12} and \cite{JR:Onireti13} for centralized and distributed MIMO systems, respectively. In \cite{JR:Heliot12}, the expression of SE, $\eta_{SE} = C_{MIMO} / W$, is generated by the MIMO ergodic capacity\footnote{For notation purpose, we consider the MIMO fading channel coefficient $\mathbf H$ with dimension $N_t \times N_r$ and denote $\mathbb{E} [\cdot]$, $(\cdot)^{H}$ to be the expectation and matrix Hermitian operation respectively. $f(\cdot)$ and $f^{-1}(\cdot)$ represent the function mapping and inverse mapping relations.}
\begin{eqnarray}
C_{MIMO} = W \cdot f(P)= W \cdot \mathbb{E}_{\mathbf H} \left[ \log_2 \det \left(\mathbf{I}_{N_r} + \frac{P}{N_t W N_0} \mathbf{H} \mathbf{H^{H}}\right) \right], \label{eqn:mimo}
\end{eqnarray}
and the ordered eigenvalue statistics of the Wishart matrix $\mathbf{H} \mathbf{H^{H}}$, where $P = f^{-1} (C_{MIMO}/W)$ $ =  f^{-1} (\eta_{SE})$ and $f(P)$ is defined as $\mathbb{E}_{\mathbf H} \left[ \log_2 \det \left(\mathbf{I}_{N_r} + \frac{P}{N_t W N_0} \mathbf{H} \mathbf{H^{H}}\right) \right]$. The SE-EE tradeoff relation can be derived via $\eta_{EE} (\eta_{SE}) = \eta_{SE} W / P = \eta_{SE} W / f^{-1} (\eta_{SE})$. For the distributed MIMO systems \cite{JR:Onireti13}, both deterministic distance dependent pathloss effects and small scale MIMO channel fading coefficient $\mathbf{H}$ are considered and the power, $P$, in the capacity expression \eqref{eqn:mimo} is weighted by the corresponding pathloss coefficients. Although the function, $f(P)$, in the distributed MIMO system has a more complicated form, we can have a uniform expression for the SE-EE tradeoff as $\eta_{EE} (\eta_{SE}) = \eta_{SE} W / f^{-1} (\eta_{SE})$. From \cite{JR:Heliot12} and \cite{JR:Onireti13}, monotonic decreasing SE-EE tradeoff relations can be obtained via theoretical analysis since $f^{-1}(\cdot)$ is a monotonically increasing function. Therefore, we cannot get EE improvement without sacrificing the SE for MIMO systems. The similar conclusion is also applicable for the DE, $\eta_{DE}$, since the additional deployment cost to install multiple antennas is not significant compared with the operational cost and can be averaged over the whole life-cycle of wireless networks. To guarantee the required data rate, $R$, or the transmission delay, $\tau$, in MIMO systems, the required power is just inversely proportional to the improvement of the SE, which turns out to be a better tradeoff for BW-PW and DL-PW.

For the MIMO tradeoff analysis under practical conditions, the circuit power consumption, $P_{c}$, should be included for idle state and proportional to the number of transmit antennas for active state.  In that case, we will have some finite value for EE even when $\eta_{SE}$ approaches to zero, which turns out to have much smaller EE for the low SE regime if compared with the ideal case. Therefore, as shown in \cite{JR:Xu131}, the SE-EE tradeoff will be a quasi-concave bell shape expression and an optimal EE value can be obtained using some efficient numerical searching algorithms. \textcolor{black}{The similar approach can be used to derive DE-EE, BW-PW, and DL-PW tradeoffs, and we plot the main fundamental green tradeoff relations in Fig. \ref{fig:fun_tra}(b) for practical MIMO systems.}

\subsubsection{Energy Efficient Solutions}
Energy efficient MIMO systems have been extensively studied recently. In the downlink, antenna precoding \cite{JR:Belmega13, JR:Varma13, JR:Hong13, JR:Jiang14,  JR:Joung13} and user scheduling \cite{CF:Liu12, CF:Zhang13} are commonly used to improve EE. For example, the optimal precoding matrix to achieve the optimal EE under the ideal MIMO channel condition has been derived in \cite{JR:Belmega13}. To be more specific, instead of using a scalar transmit power $P$, the precoding strategy, $\mathbf Q$, has been founded to maximize the MIMO channel capacity $C_{MIMO} = W \cdot \mathbb{E}_{\mathbf H} \left[ \log_2 \det \left(\mathbf{I}_{N_r} + \frac{1}{N_t W N_0} \mathbf{H} \mathbf{Q} \mathbf{H^{H}}\right) \right]$, subject to the constraint $\textrm{Tr} (\mathbf Q) \leq P$. It is showed later in \cite{JR:Belmega13} that EE follows the quasi-concavity property with respect to the precoding matrix $\mathbf Q$ and using $\mathbf Q$ to diagonalize the channel $\mathbf H$ offers the maximum capacity for a given transmit power. However, consuming all available transmit power, $P$, is not always optimal for EE maximization in downlink MIMO systems. The energy efficient precoding schemes with imperfect CSI at the transmitter have been addressed in \cite{JR:Varma13}, where the channel outage event ($C_{MIMO} < R$) has been modeled to perform the EE optimization. Compared with the precoding design in \cite{JR:Belmega13}, it is shown in \cite{JR:Varma13} that the EE optimal precoding strategy in this case should still diagonalize the estimated channel $\hat{\mathbf{H}}$ while the transmit power budget shall be a function of channel estimation errors as well. The precoding analysis has been extended to virtual MIMO systems in \cite{JR:Hong13} and \cite{JR:Jiang14} using the similar techniques,  where joint power balancing algorithms have been proposed and an open issue posted in \cite{JR:Chen11} has been addressed. The precoding matrix design based on point-to-point MIMO channel information feedback has been investigated in \cite{JR:Belmega13, JR:Varma13, JR:Hong13,JR:Jiang14}. However, the precoding schemes in multiuser MIMO scenarios need to consider multiple MIMO channels simultaneously. One of the energy efficient precoding schemes is to minimize the potential inter-user interference, such as zero-forcing based precoding as shown in \cite{JR:Joung13}. Meanwhile, to select a proper number of transmit antennas and distribute the available power among different antennas are also important to maximize the overall EE of multiuser MIMO systems. To further enhance the EE performance for the multiuser MIMO scenarios,  a greedy energy efficient user scheduling algorithm and a RF chain sleeping technique have been proposed in \cite{CF:Liu12} and \cite{CF:Zhang13}, respectively. It has been validated in system level that multiuser diversity and hardware switching can significantly improve the EE as well.

In the uplink, multiple terminals are grouped together to form uplink virtual MIMO for EE improvement \cite{JR:Onireti11, JR:Miao13, JR:Rui13}.  Mathematically, if $K$ uplink users, each with $N_t$ antennas, transmit to the same base station with $N_r$ antennas, the equivalent uplink MIMO channel $\mathbf{H}$, can be expressed as $\mathbf{H} = \mathbf{\Omega}_v \odot \mathbf{H}_v$, where $\mathbf{H}_v$ is a $N_r \times K N_t$ matrix, $ \mathbf{\Omega}_v$ is a $N_r \times K N_t$ deterministic distance dependent pathloss matrix, and $\odot$ denotes the Hadamard product. Following the similar procedures as illustrated in the downlink MIMO case, we can rewrite the MIMO capacity formula and derive the optimal EE precoding schemes for coordinated multi-point (CoMP) transmission. A generic closed-form approximation of the SE-EE tradeoff can be found in \cite{JR:Onireti11}. Through numerical simulation, it has been shown in \cite{JR:Onireti11} that uplink CoMP transmission with  optimal EE precoding is more energy efficient under the ideal and linear power consumption model. A more realistic energy efficient precoding policy considering the circuit power consumption has been developed in \cite{JR:Miao13}, where each user sets a precoding matrix to diagonalize the corresponding channel. It has been also revealed in \cite{JR:Miao13} that there exists a unique globally optimal energy efficient power allocation associated with the precoding matrix, which has a weighted water-filling form with different water levels for different users. The idea has been extended to jointly optimize the number of uplink users in the MIMO structure and the related power allocation strategy in \cite{JR:Rui13}. To reduce the searching complexity for the global optimal user set, an ergodic rate based user pre-selection approach has been proposed with marginal EE performance loss.

Apart from the aforementioned schemes, the implementation issues have been also addressed and several alternative ways for the energy efficient MIMO transmission have been proposed. For example,  antenna selection, usually regarded as a special form of power allocation strategy, e.g. to allocate zero power to the unselected antenna sets, has been analyzed in \cite{JR:Brante13} and \cite{JR:Rayel14}. Another energy efficient approach to utilize the antenna index for the information modulation called ``spatial modulation'' has been invented by Renzo {\em et. al.} \cite{JR:Renzo14}, where  the receiver side can obtain extra information bits by successfully detecting the transit antenna indices. Other energy efficient schemes including modulation diversity,  cognitive and polarization have been discussed in \cite{JR:Lee10, JR:Nicolaou10, JR:Cui12, JR:He11, JR:Chen14, JR:Kim14}, as summarized in Table \ref{table:mimo}. \textcolor{black}{In summary, the approaches in \cite{JR:Heliot12, JR:Onireti13,JR:Xu131,JR:Belmega13, JR:Varma13,JR:Hong13, JR:Jiang14} can be applied to large scale and EE oriented wireless networks with user quality of service (QoS) requirements. While the insights on realizing fairness among users can be resorted to  \cite{CF:Liu12,CF:Zhang13}. In addition, the algorithms in
 \cite{JR:Onireti11,JR:Miao13,JR:Rui13} are helpful to achieving high SE for power limited mobile users. Designing some energy efficient schemes for delay-aware applications can be in \cite{JR:Nicolaou10, JR:Cui12}.}
\begin{table}
\tabcolsep 2mm
\renewcommand{\arraystretch}{2}
\footnotesize
\caption{Taxonomy of Energy Efficient MIMO Schemes based on the Fundamental Green Tradeoffs} \label{table:mimo} \centering
\begin{tabular}{c|c|c|c|c}
\hline
\hline
Reference & Main Tradeoff & Technical Area & DL/UL & Contribution \\
\hline
\hline
\cite{JR:Heliot12, JR:Onireti13}	& SE-EE & Theoretical Analysis & DL & Characterize the theoretical tradeoff curves\\
\hline
\cite{JR:Xu131} & SE-EE & Theoretical Analysis & DL & Characterize the tradeoff curves with circuit power\\
\hline
\cite{JR:Belmega13, JR:Varma13} & SE-EE & Precoding Design & DL & Precoding with perfect/imperfect channel information\\
\hline
\cite{JR:Hong13, JR:Jiang14} & SE-EE & Precoding Design & DL & Energy efficient precoding for virtual MIMO systems\\
\hline
\cite{JR:Joung13} & SE-EE & Precoding Design & DL & Energy efficient precoding for multiuser MIMO systems\\
\hline
\cite{CF:Liu12} & SE-EE & User Scheduling & DL & Energy efficient scheduling for multiuser MIMO systems\\
\hline
\cite{CF:Zhang13} & SE-EE, DE-EE & User Scheduling & DL & Energy efficient scheduling for RF chain sleeping\\
\hline
\cite{JR:Onireti11} & SE-EE & Theoretical Analysis & UL & Characterize the theoretical tradeoff curves\\
\hline
\cite{JR:Miao13} & SE-EE & Precoding Design & UL & Energy efficient precoding with circuit power\\
\hline
\cite{JR:Rui13} & SE-EE & User Scheduling & UL & Joint uplink user selection and precoding \\
\hline
\cite{JR:Brante13, JR:Rayel14} & SE-EE & Precoding Design & DL/UL & Joint antenna selection and power control\\
\hline
\cite{JR:Renzo14} & DE-EE & Modulation Design & DL/UL & Spatial modulation for energy efficient transmission\\
\hline
\cite{JR:Lee10} & SE-EE & Modulation Design & DL/UL & Symbol rotation based modulation for MIMO diversity\\
\hline
\cite{JR:Nicolaou10, JR:Cui12} & DL-PW & User Scheduling & DL & Delay guaranteed low power MIMO transmission \\
\hline
\cite{JR:He11} & BW-PW & Resource Management & DL/UL & MIMO power minimization using cognitive spectrum\\
\hline
\cite{JR:Chen14, JR:Kim14} & SE-EE & Resource Management & DL & Energy efficient polarization and link adaptation\\
\hline
\hline
\end{tabular}
\end{table}

\subsubsection{Future Research Issues}
\begin{figure}[!t]
\centering
\includegraphics[width = 6 in]{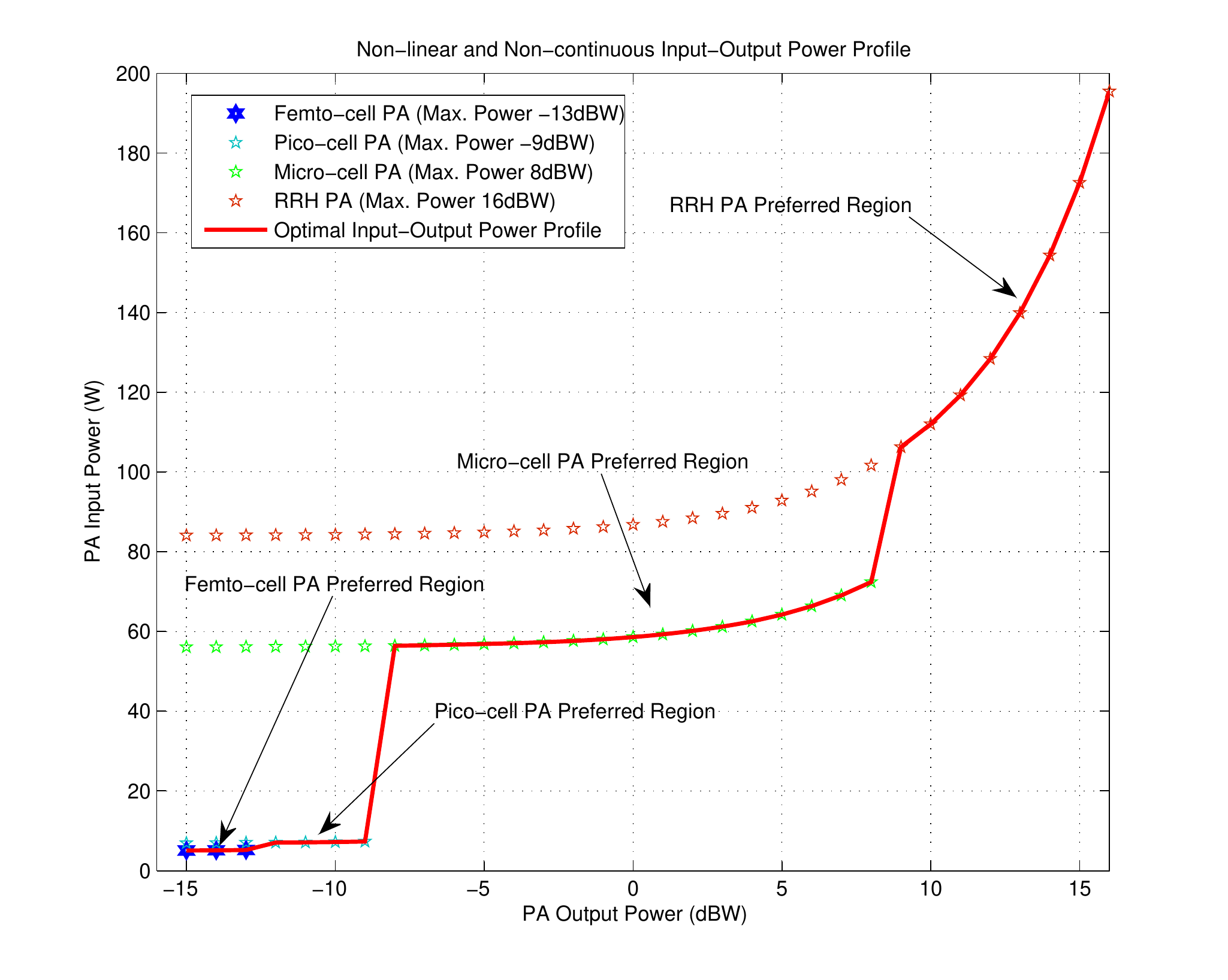}
\caption{Input and output power relations for different types of power amplifiers based on the results given by \cite{EARTH:D2.3}. In this figure, four types of PAs are considered, including femto, pico, micro, and RRH, and the red curve shows the optimal input-output power profile. As indicated in the figure, different types of PAs will have different preferred regions, e.g. -9dBW to 8dBW for micro-cell PA and 8dBW to 16dBW for RRH PA.}
\label{fig:input_output}
\end{figure}

In the current MIMO EE evaluation \cite{JR:Heliot12}, the circuit power consumption is modeled through a linearized and continuous expression, e.g. $P/{\mu_{PA}} + P_{c}$, where $\mu_{PA}$ is the efficiency of power amplifier (PA). Applying this model, the overall MIMO EE expression can be transformed into quasi-concave functions and the green tradeoff problem can be efficiently solved by standard convex relaxation/approximation techniques \cite{JR:Xu131}. Nevertheless, if we want to more accurately determine the optimal MIMO configuration of base stations or handsets in the commercial systems, we still need to consider the following two factors. One is various types of PAs with different transmission capabilities and optimal operating regions \cite{EARTH:D2.3}. For example, pico-cell PA is optimal from -12dBW to -9dBW while remote radio head (RRH) PA is efficient for more than 9dBW. With the efficient MIMO setting, the per antenna transmission power requirement will be reduced, which allows to use more efficient PAs whenever necessary. The other is that different operating regions result in different efficiencies even within the same type of PAs, which cannot be described by just a single number. Once the above two factors are combined together, the optimal input-output power relations of multi-type PAs, as shown in Fig. \ref{fig:input_output} from \cite{EARTH:D2.3}, are neither linear nor continuous. Hence, the following research issues need to be addressed with higher priority.
\begin{itemize}
\item{\em How to derive EE optimal precoding strategy with non-linear and non-continuous power model?} The main challenge is that with non-linear and non-continuous power model \cite{JR:Miao13}, the quasi-concavity property of EE no longer holds and new convex approximation method needs to be found.
\item{\em How to deploy the commercial MIMO systems with EE requirements? } With non-linear and non-continuous power model, in-depth EE analysis will be essential for the network deployment, which requires to jointly consider different PA types and dynamic PA efficiency before the cell layout stage.
\end{itemize}
\subsection{Massive MIMO} \label{subsec:m-mimo}
A M-MIMO system \cite{JR:Marzetta10} is a direct extension from traditional MIMO by installing an excessive large number of antennas at the base station. When the number of antennas is huge,  many random characteristics asymptotically turn into deterministic ones and many exact conclusions can be made. Fundamental green tradeoffs will be affected as follows.

\subsubsection{Theoretical Achievements} We can re-investigate the MIMO ergodic capacity formula \eqref{eqn:mimo} and as the number of transmit antennas $N_t$ goes to infinity, $\frac{\mathbf{H} \mathbf{H}^{H}}{N_t}$ tends to the identity matrix $\mathbf{I}_{N_r}$. In this case, the ergodic capacity for M-MIMO systems becomes,
\begin{eqnarray}
C_{M-MIMO} = W \cdot N_r \log_2 \left(1 + \frac{P}{W N_0} \right), \label{eqn:m-mimo}
\end{eqnarray}
where the randomness of channel fading has been averaged over infinite number of transmit antennas. As demonstrated in \cite{JR:Marzetta10}, without considering circuit power consumption, ten times SE improvement can be easily achieved when the number of antenna pairs exceeds one hundred\footnote{In the practical systems, we can hardly find such rich scattering environment with more than one hundred specially orthogonal channels. However, this issue can be partially compensated by deploying multiple distributed terminals, where each terminal only equips a small number of antennas.} for a given overall transmit power. The corresponding EE is growing proportionally and the resultant SE-EE tradeoff curve will be pushed outwards.  Similar conclusion applies to DE-EE tradeoff due to the limited deployment cost overhead. Following the same procedures as in the MIMO scenario, the BW-PW and DL-PW tradeoff relations can be derived accordingly.

\begin{figure}
\centering
\includegraphics[width = 6 in]{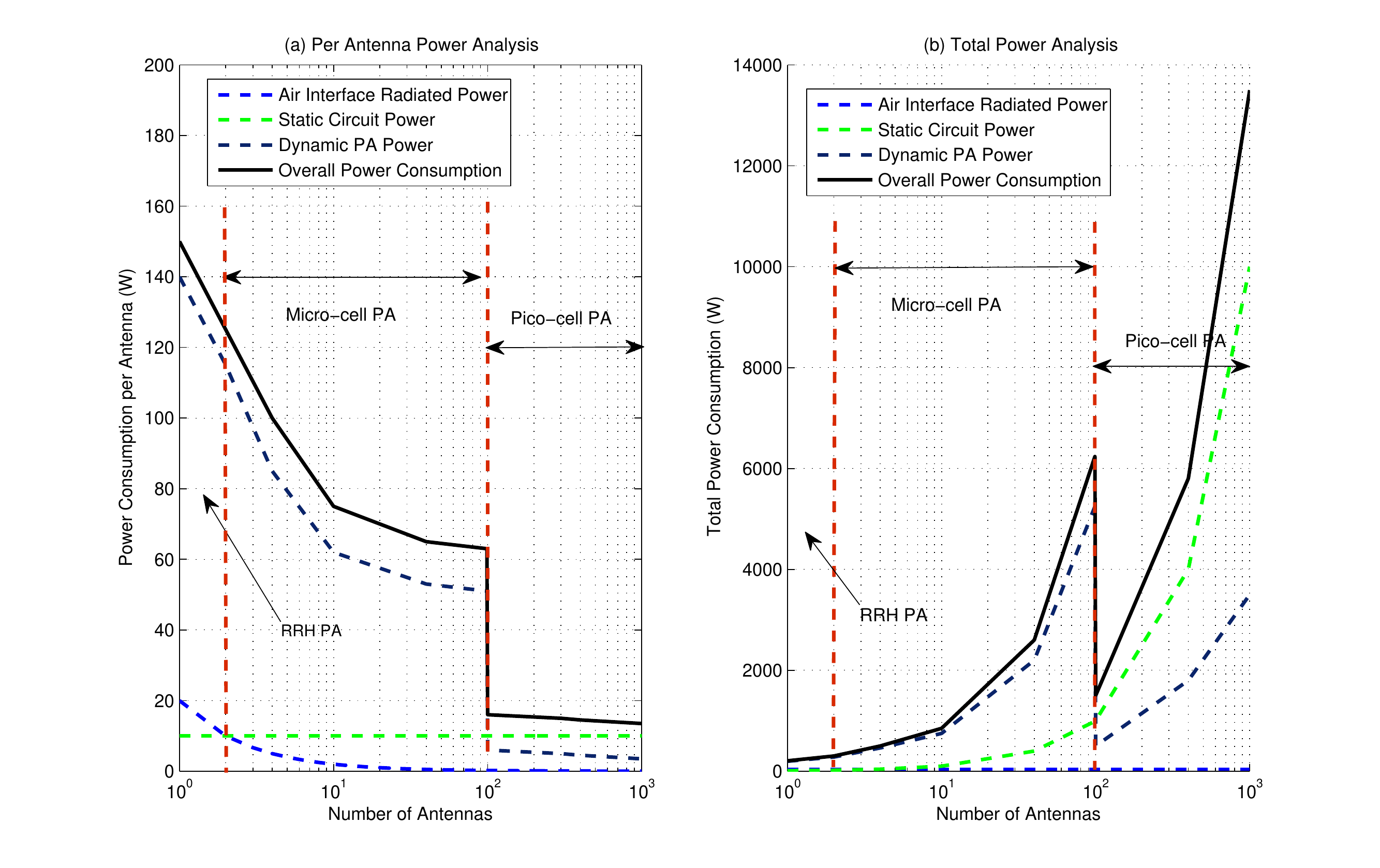}
\caption{Power consumption analysis for M-MIMO systems and the scaling effects. In this analysis, we assume the total air interface power (the sum radiated power of M-MIMO systems) is fixed to be 20W and equally distributed over all antenna elements, e.g. if we have 100 antennas, the average radiated power for each antenna is 0.2W. We compare the per antenna power and the total power consumption in sub-figure (a) and (b) respectively, where the air interface power, the static circuit power (filter, duplexer, mixer and etc.), the dynamic PA power (calculated based on\cite{EARTH:D2.3} and considered different PA types according to the radiated power requirements) and the overall power. From the analytical results, we may observe a linear power scaling relation with respect to the number of antennas.}
\label{fig:m-mimo}
\end{figure}

If we take the circuit power into consideration in the EE analysis, we will have a linear growth of static circuit power, which dominates the overall power consumption as shown in Fig. \ref{fig:m-mimo}. Therefore, we can no longer expect EE enhancement for M-MIMO systems unless the efficiency of the auxiliary circuits can be improved \cite{JR:bjornson2015optimal}. In general, the total power consumption increases although the transmit air interface power can be minimized with the M-MIMO configuration.

\begin{figure}[!t]
\centering
\includegraphics[width = 6 in]{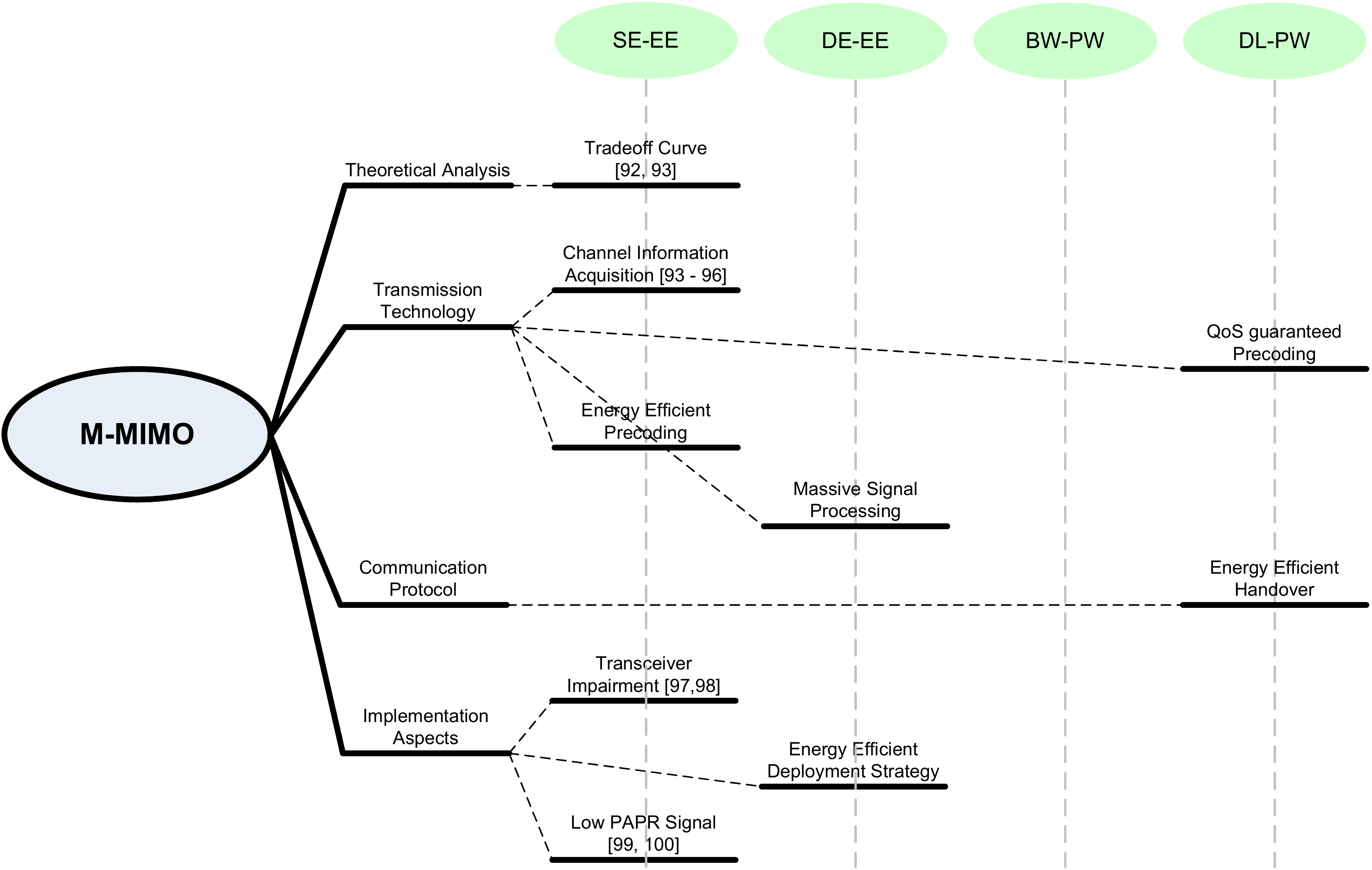}
\caption{A research roadmap for energy efficient M-MIMO technology based on the fundamental green tradeoffs.}
\label{fig:m_mimo}
\end{figure}

\subsubsection{Energy Efficient Solutions} \label{sec:ee_mmimo} To improve EE of the M-MIMO systems, we shall pay attention to several physical limitations, which may not be severe in the traditional MIMO scenario and has been ignored before. Massive channel information acquisition and the associated pilot power consumption are the first challenge since the resources for channel state acquisition is in general limited and the accuracy of the channel estimation will directly affect the achievable throughput as shown in \cite{JR:Ngo13}. In a frequency division duplexing (FDD) M-MIMO system, the required pilot resources for channel state acquisition are proportional to the number of the transmit antennas, which is unaffordable in the practical implementation. To solve this issue, a pilot beamforming scheme has been demonstrated in \cite{JR:Song14} and a semi-orthogonal pilot design approach has been proposed in \cite{CF:Zheng14}  to reduce the pilot power consumption and time-frequency resources, respectively. In a time division duplexing (TDD) M-MIMO system, channel reciprocity is applied for downlink transmission while the maximum number of supporting users is often limited by the number of orthogonal pilot sequences in the uplink direction. In order to fully utilize the M-MIMO capability, an efficient uplink pilot sequence allocation scheme has been proposed in \cite{CF:Yan15} to support more uplink transmissions. \textcolor{black}{Furthermore,  uplink and downlink pilots and transmission strategies are joint designed in \cite{JR:bjornson2015optimal} to maximize system EE. It has been shown that M-MIMO is the way to achieve high EE (tens of Mbit/Joule) in future cellular networks. However,  the pilot contamination (PC) caused by reusing pilot resources across cells affects EE significantly, which implies the necessity of actively mitigating pilot contamination in multi-cell systems.}


The second challenge lies in the transceiver hardware impairments due to the non-identical baseband and radio frequency components. Conventional MIMO systems adopt the antenna calibration approach to deal with this effect. However, with the massive antenna settings, the calibration process is usually time and energy consuming, proportional to the number of transmit antennas, and the residual distortions may accumulate to degrade the overall EE performance. A brute-force solution to address this issue is to model the hardware imperfectness as a degradation of receiving SINR and optimize the system EE based on the mismatched results \cite{JR:Bjornson14}.  A more aggressive solution has been proposed in \cite{JR:Rogalin14}, where the uplink channel measurement has been utilized for the downlink channel calibration in the general MIMO system. In order to apply this technology in the M-MIMO systems, a fast and scalable channel measuring scheme is still missing.

The last issue is that the highly dynamic output signals for the current M-MIMO transmission schemes require expensive and power inefficient RF components. This is because the average transmit power per active antenna is greatly reduced with M-MIMO configuration, yet the PAPR of the transmit signals cannot scale accordingly \cite{JR:Studer13}. To address this issue, low PAPR waveforms, such as FBMC, have been proposed for M-MIMO transmission in \cite{CF:Farhang15}.

\subsubsection{Future Research Issues}
Most of the energy efficient schemes in the regular MIMO scenario can be extended to M-MIMO systems directly if the scalability issue does not play a significant role. Besides the three physical limitations as mentioned in Section \ref{sec:ee_mmimo},  M-MIMO systems may trigger the following scaling issues that requires non-trivial solutions, as we have also demonstrated in the research roadmap in Fig. \ref{fig:m_mimo}.
\begin{itemize}
\item{\em How to design energy efficient schemes for massive signal processing?} One of the design challenges in M-MIMO systems is the huge amount of data for processing and the computational complexity for channel estimation, equalization and other detection algorithms that is generally proportional to the product of the transmit and the receive antenna numbers.  Energy efficient massive signal processing techniques, such as compressive sensing for sparse signals and distributed storage algorithms, shall be jointly considered to balance the computational delay, storage requirement, and power consumption.
   \textcolor{black}{ \item{\em How to design energy efficient pilots  for M-MIMO systems?}
Note that the pilot overhead required by conventional channel estimation schemes will be unaffordable and significantly affect the EE of M-MIMO systems. Thus, how to strike a balance between the resources spent on pilots and data symbols from the EE perspective becomes a critical problem \cite{CF:Zheng14,JR:Bjornson14}. Several possible solutions are: 1) explore the inherent sparsity of wireless channels with compressive sensing; 2) incremental pilot channel estimation design to reduce the number of pilots; 3) smarter schemes to control the pilot contamination without incurring additional energy consumption.}
\item{\em How to deploy M-MIMO systems more energy efficiently?} The energy efficient deployment problem for M-MIMO systems shall also be addressed. For example, since M-MIMO systems require strong power supply, to deploy them near the power station can greatly reduce the power dissipation in the power transmission line. Meanwhile, how to perform the frequency planning to reduce the power budget and control the inter-cell interference for M-MIMO systems is another critical direction for future research.
\end{itemize}

\section{Heterogeneous Networks}\label{sec:hetnet}
Another approach to facilitate the spatial densification is to apply heterogeneous architectures. In this section, similar to regular and massive MIMO technologies, we provide a comprehensive overview on the energy efficient solutions for HetNets and UDNs using the fundamental green tradeoffs. \textcolor{black}{To fully exploiting diversity gain and achieving high SE and EE, we focus on HetNets and UDNs with frequency reuse.}

\subsection{Heterogeneous Networks} \label{sub_sec:hetero}
A straight-forward way to address heterogeneity of wireless communications is to use macro cells to provide the coverage and small cells to improve the hot-spot throughput. This technology is typically called HetNets. Fundamental green tradeoff analysis for HetNet has been addressed before and can be summarized as follows.

\subsubsection{Theoretical Achievements} Without loss of generality, the ergodic capacity of the HetNet can be written as,
\begin{eqnarray}
C_{HetNet} = W \cdot \mathbb{E}_{\{h_i\}} \left[\max_{i \in \{1, 2, \ldots, N_c\}}\log_2\left(1 + \frac{P_i \|h_i\|^2}{W N_0 + \sum_{i'=1, i'\neq i}^{N_c} P_{i'} \|h_{i'}\|^2}\right)\right], \label{eqn:HetNet}
\end{eqnarray}
\textcolor{black}{where $N_c$ is the number of cells and $P_i$ is the transmit power of cell $i$. $h_i$ is the fading coefficient between the user and the $i^{th}$ cell.} Due to the HetNet architecture, we may have peak throughput within small cells that incurs heterogeneous capacity layout for the network, and {\em area SE (ASE)} \cite{JR:Alouini99, JR:Chandhar14, JR:Shakir14,CF:wang2010energy,CF:lorincz2012bit} rather than SE is applied as the performance measure. ASE addresses the fairness issues between the macro cells and the small cells by averaging over the heterogeneous cell coverage and can be mathematically expressed as
\begin{eqnarray} \label{eqn:ASE}
\eta_{ASE} = \max_{P_i \geq 0, \ \sum_{i} P_i = P} \frac{\int_{\mathcal{A}} \eta_{SE} (\{P_i\}, \textrm{d} \mathcal{A}) \cdot \textrm{d} \mathcal{A}}{|\mathcal{A}|}, \label{eqn:ase}
\end{eqnarray}
where $\mathcal{A}$ denotes the interested heterogeneous coverage area and $|\mathcal{A}|$ is the area of $\mathcal{A}$. $\eta_{SE} (\{P_i\}, \textrm{d}\mathcal{A}) = C_{HetNet} / W$ is the SE achieved within the area of $\textrm{d} \mathcal A$ for a given transmit power set, $\{P_i\}$. Similarly, area EE (AEE) can be defined as the ratio of the ASE and the total power consumption, $P$, i.e. $\eta_{AEE} = \eta_{ASE} / P$ \cite{CF:Lorincz12}. For the SE-EE tradeoff characterization, the power consumption, $P$, can be written as a function of $\eta_{ASE}$ through equation \eqref{eqn:ase}.

HetNet provides high throughput in hot-spot areas with marginal power resource, and therefore, results in high ASE and AEE compared to the traditional homogeneous networks, as verified in \cite{CF:He13} and \cite{JR:Kim14_W}. The same method can be applied for DE-EE, BW-PW and DL-PW relations as well. The ASE improvement reveals better tradeoffs and lower power consumption in the ideal case. In the practical environment, in addition to the circuit power consumption as mentioned in many other scenarios, the deployment of small cells also incurs extra operational expenditure for site maintenance, which reduces the DE performance as shown in \cite{CF:He12}.  Meanwhile, HetNet provides the flexibility to allocate orthogonal or non-orthogonal frequency bands for different cells and eventually improves the BW-PW tradeoff since inter-cell interference can be controlled via proper frequency planning as explained in \cite{CF:Zhang12} and \cite{CF:Zhang131}. It is also worth to mention that the delay performance for HetNet involves heterogeneous queueing from macro and small cells, which is completely different from the traditional homogeneous networks. The detailed DL-PW tradeoff analysis for HetNet can be found in \cite{JR:Kong13}. \textcolor{black}{As a result, four fundamental tradeoff relations can be characterized by Fig. \ref{fig:fun_tra}(c), where EE drops and PW raises in the low SE/DE and BW/DL regimes}\footnote{To be more precise, the performance behaviors of HetNets are highly depending on the actual user distributions, mobilities, traffic variations and other factors. In this paper, we focus on the statistical results and ignore the instantaneous traffic dynamics for illustration purpose.} .

\subsubsection{Energy Efficient Solutions}
To improve the EE performance for HetNet and resolve open issues raised in \cite{JR:Chen11}, various types of energy efficient solutions have been proposed according to different time scales. Based on the operational time periods, we can categorize them into the following three levels.
\begin{enumerate}
\item{\em Short Time Scale - Resource Management via Scheduling \& Load Balancing:}
In the short time scale (e.g. seconds or minutes), one of the unique problems in heterogeneous resource management is intra-cell interference among macro cells and small cells, where the frequency (sub-channel/subcarrier) and power resources need to be balanced based on the rate or QoS requirements. To fulfill the goal, the brute-forth solution is to model intra-cell interference into the original radio resource management problem and then perform resultant hybrid resource management \cite{JR:Mclaughlin11, JR:Pramudito13}. For example, the intra-cell interference control for a two-tier HetNet using power management has been discussed in \cite{JR:Mclaughlin11}, where  the small cell transmission power level is adjusted with respect to the received power level from the nearest macro cell. In \cite{JR:Pramudito13}, a hybrid approach has been proposed to jointly allocate the frequency and power resources, where the frequency partitioning scheme is utilized through fractional frequency reuse among macro and small cells. An alternative way is to turn the harmful inter-cell interference into desired signal via macro-small cells cooperation. Coordinated multi-point (CoMP) precoding has been proposed in \cite{JR:Xu14} and \cite{JR:He14} as one of the most efficient solutions for cell cooperation to improve HetNet EE, where the zero-forcing technique to control the interference is used in \cite{JR:Xu14} and the multi-cell joint transmission is utilized in \cite{JR:He14} to maximize the weighted sum EE. Another approach for cell cooperation is through dynamic user association and load-aware cell migration operation \cite{JR:qing16_std}. The corresponding analytical framework has been developed for single radio access technology (RAT) in \cite{JR:Son11} and \cite{JR:Ismail14} and for multi-RAT in \cite{JR:Lim14}, respectively. Besides the above cell domain cooperation, opportunistic user domain cooperation has been proposed for user load balancing in \cite{JR:Wang14} as a complementary solution to improve the HetNet EE.

\item{\em Medium Time Scale - Topology Management via Cell Activation/De-Activation:}
In the medium time scale (e.g. hours or days), HetNet architecture has a unique advantage to activate/de-activate small cells to optimize EE according to the traffic dynamics since the coverage issues can be well solved by macro cells. In \cite{JR:Soh13}, an EE optimization framework has been developed for $K$-tier HetNets. Based on that, an energy efficient cell sleeping policy to minimize network power consumption has been developed and optimized. In \cite{JR:Conte11}, an enhanced cell activation/de-activation strategy has been proposed to address the capacity gap by gradually adjusting the transmit power in order to prevent from the call-drop events. Nevertheless, in commercial networks, the transition for mobile users to migrate among different cells is usually time-consuming and the signal exchange overhead is significant as well. Fortunately, the above strategy is usually performed in the low traffic period, such as in mid-night.

\item{\em Long Time Scale - Network Management via Cell Deployment: }
In the long time scale (e.g. months or years), a meaningful question becomes how we shall deploy HetNet more energy efficiently. As a starting point, a special topology has been considered in \cite{JR:Zeeshan13}, where all small cells are located around the reference macro cell, to improve the cell edge throughput and system EE. By utilizing a simple uplink power adaptation scheme, 30\% power can be reduced compared with macro-only networks. The topology constraint is relaxed to arbitrarily deploy small cells for given macro cell locations in \cite{JR:Coskun14}. The proposed energy efficient deployment strategy is able to improve EE by 40\% to 65\%. In \cite{JR:Cao13, JR:Peng15, JR:Cao131}, joint macro and small cell energy efficient deployment has been investigated to minimize the overall HetNet power consumption. In particular, the macro and small cell density adjustment has been proposed in \cite{JR:Cao13} for two-tier HetNet and its impacts on the network coverage performance has been evaluated in \cite{JR:Peng15}. In \cite{JR:Cao131}, the energy efficient deployment strategy has been extended to jointly consider the partial spectrum reuse. With practical parameter settings, over 50\% network power consumption can be saved. \textcolor{black}{In \cite{CF:ghazzai2014optimized,CF:yaacoub2014lte,CF:yaacoub2014lte1}, some probabilistic techniques, such as meta-heuristic
algorithm and simulated annealing, have been also exploited to address the cell deployment issue. Specifically, the numbers of required BSs and their locations are optimized in \cite{CF:yaacoub2014lte} by using simulated annealing under different user distribution scenarios. Then, similar problem is investigated in \cite{CF:yaacoub2014lte1} in the presence of fixed open femtocell access points (FAPs). }
\end{enumerate}

Moreover, the above energy efficient schemes can be combined together to provide better tradeoffs. For example, cell de-activation needs to be integrated with load balancing for the active calls \cite{JR:Cho13}. The deployment of HetNets shall also consider the instantaneous energy dynamics if renewable resources are applied \cite{JR:Piro13}. Another interesting remark is that long time scale user information can also help improve the short time scale adaptation from the recent reports \cite{JR:Zhang141} and \cite{JR:Zhang14}, e.g. the historical user behaviors allow the network to opportunistically schedule the coordinated multicast transmission for EE enhancement. We refer readers to Table \ref{table:hetnet} for an overall comparison.
\begin{table}
\tabcolsep 2mm
\renewcommand{\arraystretch}{2}
\footnotesize
\caption{Taxonomy of Energy Efficient HetNet Schemes based on the Fundamental Green Tradeoffs} \label{table:hetnet} \centering
\begin{tabular}{c|c|c|c|c}
\hline
\hline
Ref. & Main Tradeoff & Technical Area & DL/UL & Contribution \\
\hline
\hline
\cite{CF:He13, JR:Kim14_W} & SE-EE & Theoretical Analysis & DL & Characterize the theoretical tradeoff curves\\
\hline
\cite{CF:He12} & DE-EE & Theoretical Analysis & DL & Characterize the tradeoff curves with circuit power\\
\hline
\cite{CF:Zhang12, CF:Zhang131} & BW-PW & Theoretical Analysis & DL & Characterize the tradeoff curves with circuit power\\
\hline
\cite{JR:Kong13} & DL-PW & Theoretical Analysis & DL & Characterize the tradeoff curves with circuit power\\
\hline
\cite{JR:Mclaughlin11, JR:Pramudito13} & SE-EE & Resource Management & DL & hybrid resource management with intra-cell interference\\
\hline
\cite{JR:Xu14, JR:He14} & SE-EE & Precoding Design & DL & Coordinated multi-point precoding for cell cooperation\\
\hline
\cite{JR:Son11, JR:Ismail14, JR:Lim14}  & SE-EE, DE-EE & User Scheduling & DL & Dynamic user association and load balancing for cell cooperation\\
\hline
\cite{JR:Wang14} & SE-EE & User Scheduling & DL/UL & Opportunistic user cooperation for user load balancing\\
\hline
\cite{JR:Soh13, JR:Conte11} & SE-EE, DE-EE & Resource Management & DL & Energy efficient cell sleeping policy for cost minimization\\
\hline
\cite{JR:Zeeshan13, JR:Coskun14, JR:Cao13} & DE-EE & Network Deployment & DL/UL & Energy efficient deployment strategy for power reduction \\
\hline
\cite{JR:Cao131} & BW-PW & Network Deployment & DL &  Energy efficient deployment with partial spectrum reuse \\
\hline
\cite{JR:Peng15} & DE-EE & Network Deployment & DL/UL & Joint macro-small cell deployment strategy for coverage\\
\hline
\cite{JR:Cho13} & SE-EE, DE-EE & Resource Management & DL & Hybrid load balancing and cell de-activation scheme\\
\hline
\cite{JR:Piro13} & DE-EE & Network Deployment & DL & Efficient deployment strategy with renewable energy\\
\hline
\cite{JR:Zhang141, JR:Zhang14} & SE-EE & User Scheduling & DL & Coordinated multicast transmission using historical information\\
\hline
\hline
\end{tabular}
\end{table}
\textcolor{black}{In summary, \cite{CF:He13, JR:Kim14_W,JR:Mclaughlin11, JR:Pramudito13,JR:Xu14, JR:He14} have provided EE-SE regions as well as specific resource allocation algorithms for HetNets.  The results in \cite{CF:Zhang12, CF:Zhang131,JR:Cao131} shed lights on how to strike a balance between bandwidth and power while meeting the users data rate requirements. In addition, some useful guidances on green network deployment with or without BS switching can be found in \cite{JR:Zeeshan13, JR:Coskun14, JR:Cao13} and an effective approach to exploiting renewable energy to realize green HetNets has been provided in \cite{JR:Piro13}. Furthermore,  a method to identify the relationship between power consumption and packet delay has been developed in \cite{JR:Kong13}. }
\subsubsection{Future Research Issues}
\begin{figure}
\centering
\includegraphics[width = 6 in]{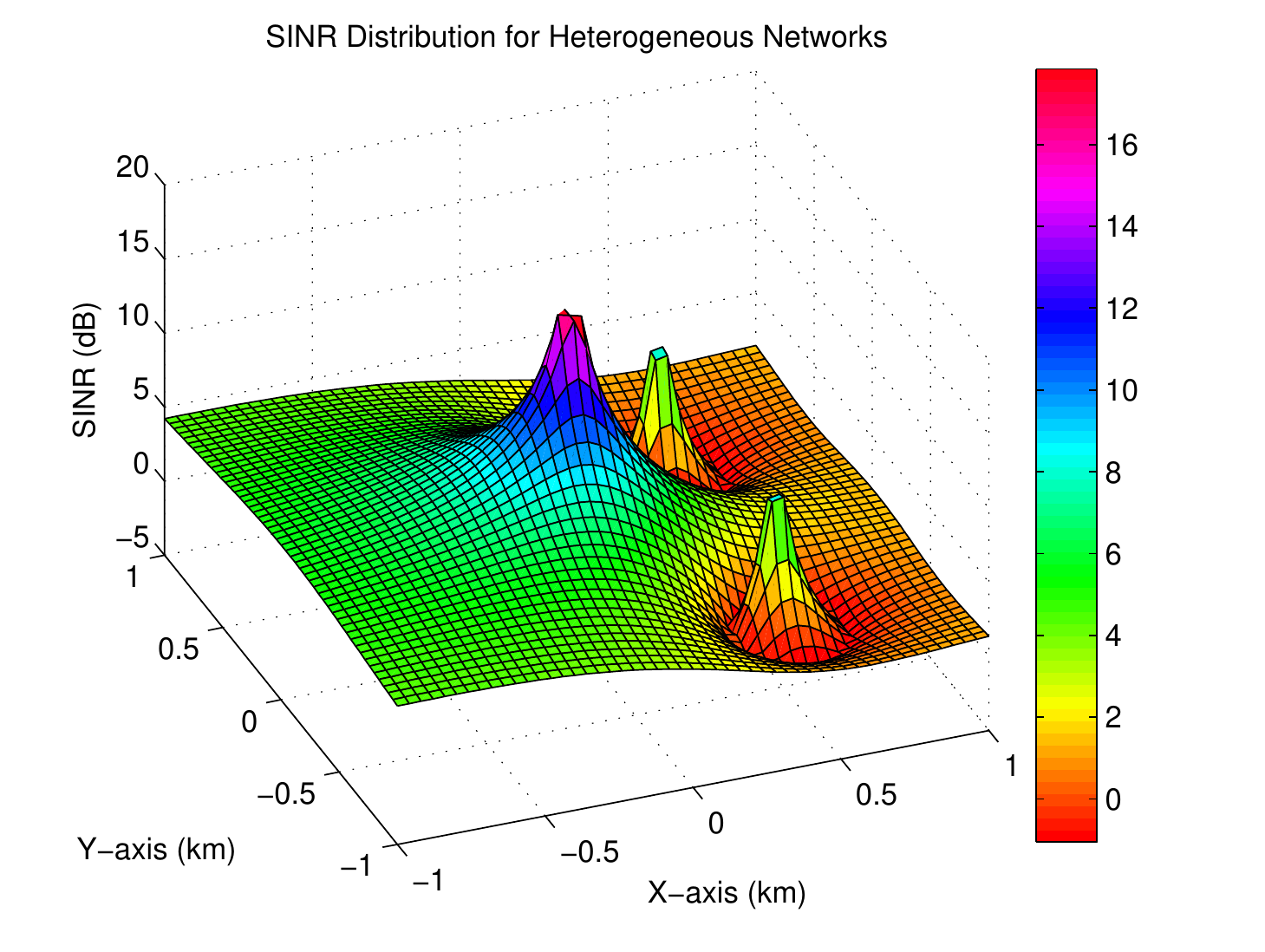}
\caption{SINR distribution with the HetNet architecture. In this figure, the macro base station is located in the origin, and two small cells are located in $(0.5, \pm 0.5)$ respectively. We applied the single frequency reuse strategy for the network, and as shown in the figure, the SINR function may become non-continuous at the cell boundary.}
\label{fig:SINR}
\end{figure}

The EE evaluation for HetNets, as we have mentioned in the previous part, relies on the statistical geometry theory to calculate the average SINR and derive other green performance measures thereafter. However, there are only limited types of base stations and limited locations to deploy in reality and the common assumption to achieve fully randomized cell distribution and continuous SINR expression may fail. For example, if we consider one reference macro cell with two small cells as shown in Fig. \ref{fig:SINR}, the continuity property of SINR function in the randomized topology no longer holds\footnote{In the randomized topology, the SINR of the mobile user at a random distance can be modeled by a continuous function based on the stochastic geometry theory \cite{JR:Andrews11}. However, in the practical network deployment with limited base station types and locations, we may have some SINR gaps at the cell boundary as shown in Fig. \ref{fig:SINR}, which cannot be described by a simple continuous function as in the random topology case.} and the EE optimal deployment problem becomes challenging. Another important issue is that the control signaling delivery and exchange are usually underestimated by wireless researchers. For instance, due to the requirements from user cell-selection/re-selection/handover, static system information broadcasting shall be performed by every cell in the heterogeneous coverage, which prevents from timely de-activation for energy saving. Hence, the following research directions are recommended for more investigation.
\begin{itemize}
\item{\em How to design low complexity EE optimal deployment schemes with limited base station types and locations?} In order to develop an analytical framework for EE optimal deployment, we can use the mixed integer programming \cite{BK:Nemhauser} to model the limited base station types and locations. However, the optimal solution requires exhaustive search over all possible combinations of base station types and locations, which is in general complexity prohibited. Low complexity algorithms, such as branch-and-bound or branch-and-cut \cite{BK:Nemhauser}, are useful directions to develop EE optimal deployment schemes, which require more dedicated research efforts.
\item{\em How to control the signaling overhead and facilitate timely cell activation/de-activation?} No-cell concept with cloud radio access network (C-RAN) architecture \cite{JR:I14} and the functionality separation scheme \cite{JR:Xu13} have just been proposed to address this issue. However, the current solutions require perfect synchronization and complicated signaling exchange between macro cells (for coverage) and small cells (for throughput enhancement), which presents from large scale deployment of this technology. Meanwhile, how to connect with higher layer separation technologies, such as network functionality virtualization (NFV) \cite{NFV:12} and software defined network (SDN) \cite{JR:Goth11}, is still open and requires further investigation.
\end{itemize}

\subsection{Ultra Dense Network}
Similar to M-MIMO, UDNs \cite{JR:Yunas15} can be regarded as a ``massive'' version of HetNets, where small cells are densely deployed over hot-spot areas. With the UDN environment, the coverage of each small cell is relatively small and the inter-cell interference with dense neighboring cells is whitened, which makes the research on the fundamental green tradeoffs easier.

\subsubsection{Theoretical Achievements}
By the ultra dense deployment assumption of small cells, users in each small cell will experience a flat SINR environment and the ergodic capacity formula for a UDN can be simplified from equation \eqref{eqn:HetNet} as
\begin{eqnarray}
C_{UDN} = W \cdot \mathbb{E}_{\{h_i\}} \left[\max_{i \in \{2, \ldots, N_c\}} \log_2\left(1 + \frac{P_i \|h_i\|^2}{W N_0 + P_1 \|h_1\|^2 + \mu(P-P_1-P_i)}\right)\right], \label{eqn:UDN}
\end{eqnarray}
\textcolor{black}{where $N_c$ is the number of small cells and $P$ is the total transmit power of the macro cell and small cells. For the illustration purpose, the individual transmit powers of the macro cell and the small cell $i$ are denoted as $P_1$ and $P_i$, respectively. The whitening coefficient  denoted as $\mu$ jointly considers the whitened inter-cell interference of dense neighboring cells and interference cancellation effect\footnote{In the ideal case, if UDN network can perfectly cancel the interference from other neighboring cells, $\mu = 0$ and the ergodic capacity of UDN becomes the sum capacity of each individual small cell.}. }In \eqref{eqn:UDN}, we ignore the capacity contribution from the macro cell for brevity\footnote{In the UDN scenario, the ergodic capacity contributed by the macro cell is marginal due to the much deeper pathloss fading if compared with small cells.}. Similarly, ASE as expressed in \eqref{eqn:ASE}, can be simplified as,
\begin{eqnarray}
\eta_{ASE}  = \max_{P_i \geq 0, \ \sum_{i =1}^{N_c} P_i = P} \frac{\sum_{i = 2}^{N_c} \eta_{SE} \left(\{P_i\}, \mathcal{A}_i \right) \cdot |\mathcal{A}_i |}{|\mathcal{A} |}.
\end{eqnarray}
If we further consider the symmetric assumption for small cells, e.g. $|\mathcal{A}_i |$ and the distribution of $h_i$ are identical for $i \geq 2$, the above maximization can be achieved when all the small cells transmit the same power, i.e., $P_i = (P - P_1)/(N_c - 1)$ for $i \geq 2$. In that case,
\begin{eqnarray}
\eta_{ASE} = \max_{P_1 \geq 0} \ \frac{\eta_{SE} (P_1) \cdot \left(N_c - 1\right) |\mathcal{A}_i |}{|\mathcal{A}|},
\end{eqnarray}
where $\eta_{SE} (P_1) = C_{UDN}/W $ can be determined through
\begin{eqnarray}
\eta_{SE} (P_1) =  \mathbb{E}_{h_i} \left[ \log_2 \left(1 + \frac{(P-P_1) \|h_i\|^2}{(N_c - 1)(W N_0 + P_1 \|h_1\|^2) + \mu (N_c - 2) (P - P_1)}\right) \right].
\end{eqnarray}

AEE analysis of a UDN follows the same procedure as the HetNet scenario and from which the SE-EE tradeoff relation can be derived. Regarding DE-EE tradeoff, a UDN involves additional deployment costs on site renting and infrastructure maintenance, which induces DE drop in the high cell density scenario when ASE is saturated \cite{JR:Tombaz11}.

The BW-PW and DL-PW tradeoffs will be significantly improved due to its flexible frequency reuse and distributive protocol management capability. With the circuit power considered, proportional expending of the total network power consumption for a UDN will not be surprising as illustrated in \cite{JR:Yunas15}. The overall network EE is saturated even if the network ASE is scaling up. For power related tradeoffs, due to the circuit power overhead of massively deployed small cells, we will have an enhancement in the total power consumption of a UDN, which is more or less similar to the M-MIMO scenario.

\subsubsection{Energy Efficient Solutions}
To deliver an energy efficient UDN, one of the major concerns is mobility management when fast moving entities passing through a UDN region. As a legacy approach, frequent handover process will be triggered in this case, and the resultant signaling overhead could be significant \cite{JR:Xu13}. To overcome this obstacle, a functionality separation scheme has been proposed in \cite{JR:Xu13} to isolate the control process so that the mobility issues are within the macro cells and the UDN is only utilized for throughput boosting. A more aggressive approach is based on a user-centric design to visualize the control-data decoupling, e.g. users could enjoy uplink/downlink transmission through user-preferred cells in the UDN environment rather than scheduled cells in the conventional systems. The tightly coupled user-cell relationship in this case is therefore detached. This is so-called ``no-cell concept'' in \cite{JR:I14} or hyper cellular network in \cite{JR:Liu14}.

Another major issue is large scale cooperation among massive UDN nodes, which includes the challenges of massive information exchange and processing. Distributed interference management based on local information coordination is a possible solution to minimize the information exchange. For example, the range expansion and interference coordination has been proposed in \cite{JR:Guvenc11} to balance the capacity distribution of massive small cells and the distributed power control mechanism has been proposed in \cite{JR:Foschini93} for heterogeneous cell coverage and in \cite{JR:Rasti10} and \cite{JR:Rasti11} for dynamic target SIR tracking. Joint cell association and distributed power control scheme for interference minimization that has been originally proposed in \cite{JR:Yates95} for code division multiple access (CDMA) systems has been extended to an energy efficient UDN in \cite{JR:Hossain14}. Another choice is to deal with ultra dense cooperative signal processing using central cloud \cite{JR:I14}, where the strict round-trip delay requirement over front-haul links is likely to be the performance bottleneck. In \cite{JR:Nakamura13}, the detailed delay requirement and potential issues for front-haul processing in LTE advanced HetNet have been analyzed and a joint design of baseband unit processing and front-haul transmission for a UDN has been proposed in \cite{JR:Wang_R14}. A holistic view of the above two solutions for large scale cooperation in the UDN can be found in \cite{JR:Hwang13}.

\subsubsection{Future Research Issues} 
\begin{figure}
\centering
\includegraphics[width = 6 in]{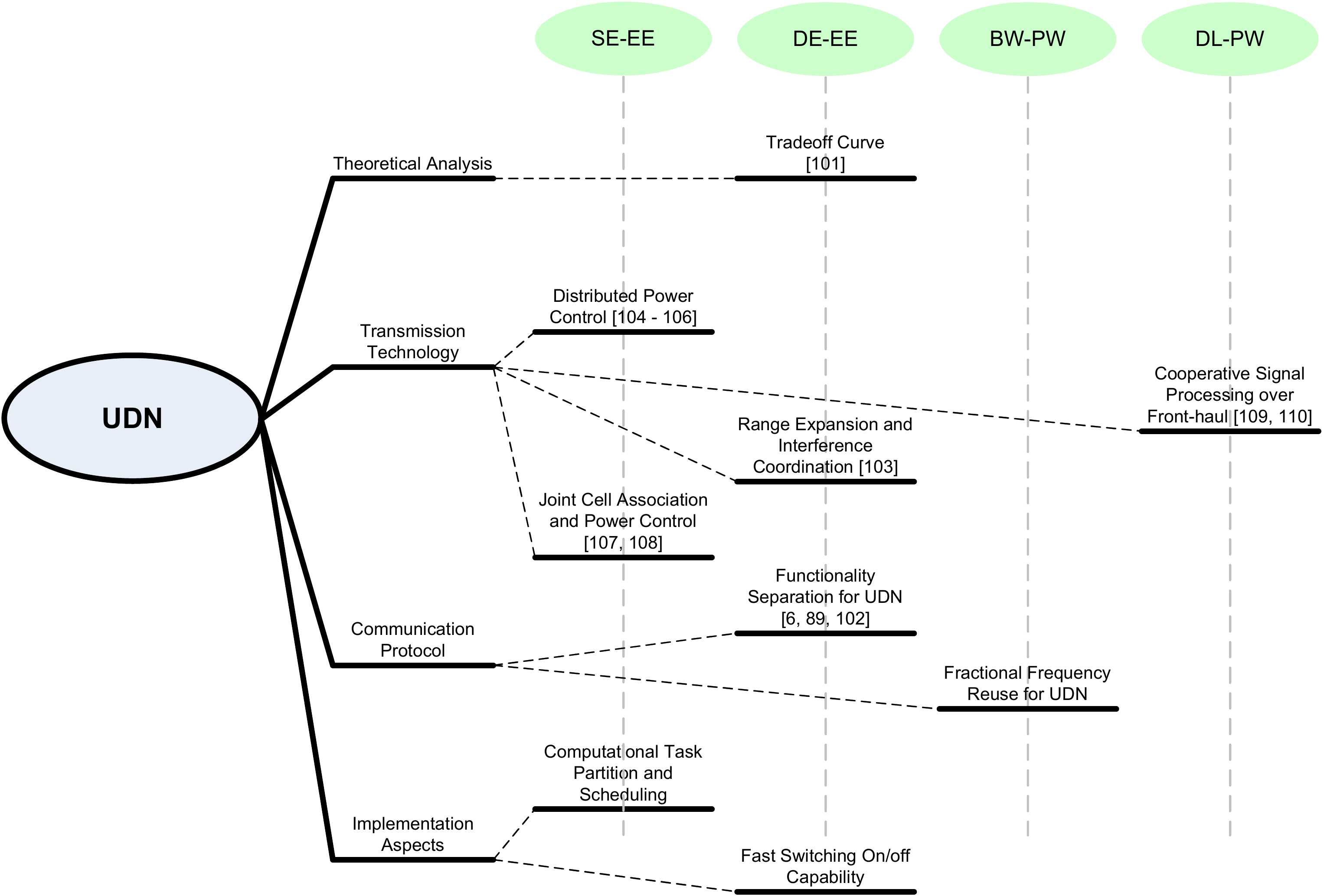}
\caption{A research roadmap for energy efficient UDN technology based on the fundamental green tradeoffs.}
\label{fig:UDN}
\end{figure}

Similar to M-MIMO systems, we can apply most of the energy efficient HetNet schemes to UDNs directly. In addition to the mobility management and large scale cooperation issues as mentioned before,  the following issues are also important for energy efficient UDN design as summarized in Fig. \ref{fig:UDN}.
\begin{itemize}
\item{\em How to partition and schedule the computational tasks among UDN nodes?} Due to the massive deployment of UDN nodes, we cannot afford expensive hardware for every node. An economic and energy efficient approach is to distribute the computational tasks over the heterogeneous UDN nodes with different computational capabilities \cite{JR:Wang_R14}. Thus, an efficient and delay guaranteed computation partition and scheduling mechanism is desired.
\item{\em How to design energy efficient implementation schemes with fast switching on/off capabilities?} Another implementation technique for energy efficient UDN is to provide fast switching on/off capabilities. In the 5G scenario, wireless traffic profile and best matching UDN architecture are highly dynamic while the legacy cell activation/de-activation scheme for energy efficient HetNet with medium time scale is unable to support the instantaneous adjustment \cite{JR:Hwang13}. Advanced hardware technologies and software system control protocols to provide fast and frequent UDN node switching on/off capabilities will be a promising direction.
\end{itemize}

\section{Other Technologies} \label{sec:mmWave}
In the above, we have discussed EE related issues on some important technologies, including OFDM and NOA, regular and massive MIMO, HetNets and UDNs.  \textcolor{black}{In this section, we will briefly introduce EE for other techniques, including millimeter wave (mmWave) communications and cognitive radios (CRs). Due to page limit, we are unable to include all of advanced technologies, such as interference alignment, unlicensed Long Term Evolution (LTE), and wireless power transfer. We refer interested readers to \cite{JR:zhao2015adaptive,JR:yoon2014energy,JR:tang2015energy,JR:abinader2014enabling,JR:zhang2015coexistence,zhangrui13_mimo,
JR:qing15_wpcn_twc,CF:qing15_wpcn,NgLS14,ng2015secure} for more information.}

MmWave communications \cite{JR:Rappaport13} are promising for 5G networks, where the typical frequency bands include 15GHz \cite{5g:14}  and 28/38GHz  \cite{JR:Rappaport13}. With the help of massively aggregated mmWave spectra, mmWave communications are able to provide significant throughput gain over the traditional cellular systems. Due to low price of mmWave devices, the total deployment cost of mmWave technology is expected to be sustainable, which makes it an attractive solution for cost efficient throughput improvement. However, existing research results for mmWave focus on the feasibility analysis \cite{JR:Dehos14, JR:Roh14, JR:Wei14} and the statistical channel models \cite{JR:Rappaport131, JR:Sulyman14} while energy efficient mmWave communications are still in the preliminary stage. One possible solution is to reduce the power consumption of mmWave systems by sharing the same RF components and antennas for multiple mmWave bands, and as illustrated in \cite{JR:Dehos14}, the overall power consumption for multiple mmWave bands will {\em not} scale up as the M-MIMO or UDN case. Another energy efficient mmWave design is to use directional antenna beams rather than energy inefficient omni-directional solutions, where the user mobility may become an issue. A candidate solution is to invent fast and accurate beam tracking or selection algorithms in \cite{JR:Huang10} to avoid unnecessary waste of radio powers. In addition, the modulated signal waveform will greatly affect the EE of mmWave systems. For example, if OFDM-type of modulation is still used, the mmWave amplifier still suffers from high PAPR and leads to harmful out of band emission. Therefore, investigating alternative approaches, such as single-carrier schemes \cite{JR:Kato09}, will be worthwhile. \textcolor{black}{Furthermore,  the power consumption model should be revisited to characterize the new features of the mmWave hardware. Besides, MmWave communications can be combined with other technologies to achieve better EE performance. One proposal \cite{JR:Hong141, JR:Han15} is to jointly consider mmWave and M-MIMO systems since the required inter-antenna spacing shrinks in mmWave bands and allows to integrate a lot more antenna elements and form narrow beams to extend the coverage. A hybrid solution to incorporate mmWave and UDN schemes is another possible way to deliver energy efficient networks \cite{JR:Ghosh14, JR:Mehrpouyan15, JR:Zheng15}, where macro cells working at the low frequency band provide the coverage layer and small cells working at mmWave bands contribute as the rate boosting layer.}

CR is another candidate technology for 5G communications. The SE-EE tradeoff for CR networks has been analyzed in \cite{JR:Hong14} and other EE related tradeoffs have been summarized in \cite{JR:Eryigit14}. Based on the fundamental green tradeoffs in CR networks, an energy efficient design framework has been discussed in \cite{JR:Jiang141}, which includes spectrum sensing, spectrum sharing, and network deployment schemes. In particular, the energy efficient spectrum sensing mechanisms have been studied in \cite{JR:Eryigit13} and \cite{JR:Sun13} to address the heterogeneous channel conditions. In \cite{JR:Mesodiakaki14} and \cite{JR:Treust13}, user association and power control strategies for energy efficient spectrum sharing have been discussed. In \cite{JR:Liu13}, an energy efficient deployment scheme has been designed, which jointly considers dynamic interference management and network coordination. {By incorporating CR technology in the HetNet environment, the cognitive spectrum can be efficiently shared between macro cells and small cells, and an energy efficient resource allocation algorithm to fully exploit the cognitive capability has been proposed in \cite{JR:Xie12}. The cognitive idea has also been extended to 5G networks, where we refer readers to \cite{JR:Gupta15} for more discussion.}

\section{Centralized or Distributed?} \label{sec:hetero}
For future 5G communications, throughput, delay, connectivity, and other requirements are all scaling with the order of 100 to 1,000 times while the energy consumption is not allowed to scale up at the same speed. In this case, the EE target is becoming more critical for the technology evolution than ever before. As characterized in Table~\ref{table:5g} however, the candidate 5G technologies fail to achieve proportional EE enhancement for large scale systems. The existing isolated energy efficient schemes,  as we have analyzed in the previous sections, may conflict with each other, resulted in diminished EE gain. For example, energy efficient new waveform design, if combined with M-MIMO or UDN systems, requires non-linear inter-antenna interference cancellation \cite{JR:Caus14, JR:Michailow14, JR:Andrews14} and is complexity and power hungry. Moreover, 5G communications target to support diversified applications with hundreds times of delay and connectivity variations\footnote{For example, in the vehicle-to-vehicle applications, the round-trip delay for communication systems shall be controlled within 1 millisecond while in the smart metering applications, second level delay can be tolerated \cite{METIS:12}.} and energy efficient schemes for 5G networks require a careful top-down design to fulfill the heterogeneous applications and requirements by exploiting infrastructure, protocol, and implementation level heterogeneities. Therefore, whether to use centralized or distributed techniques and how to combine them become very interesting.
\begin{table}
\tabcolsep 2mm
\renewcommand{\arraystretch}{1.6}
\footnotesize
\caption{Fundamental Green Tradeoffs for Key 5G Technologies} \label{table:5g} \centering
\begin{tabular}{c||c|c}
\hline
\hline
Tradeoffs & SE-EE / DE-EE & BW-PW / DL-PW \\
\hline
\hline
               	& SE: Increase with the number of users & BW: Decrease with the number of users \\
		& and become saturated for large scale case & and become saturated for large scale case \\
NOA		& DE: Increase for small number of users & DL: Decrease with the number of users\\
		& and decrease for large number of users & and become saturated for large scale case\\
		& EE: Increase for small number of users & PW: Decrease for small number of users\\
        	       	& and decrease for large number of users & and increase for large number of users\\
\hline
               	& SE: Increase with the number of antennas & BW: Decrease with the number of antennas \\
               	& and become saturated for large scale case & and become saturated for large scale case\\
M-MIMO	& DE: Increase for small number of antennas & DL: Decrease with the number of antennas \\
		& and decrease for large number of antennas & and become saturated for large scale case \\
        	       	& EE: Increase for small number of antennas & PW: Decrease for small number of antennas\\
               	& and decrease for large number of antennas &  and increase for large number of antennas\\
\hline
               	& SE: Increase with the number of small cells & BW: Decrease with the number of small cells \\
		& DE: Increase with the number of small cells & DL: Decrease with the number of small cells \\
UDN		& and become saturated for large scale case & and become saturated for large scale case\\
        	       	& EE: Increase with the number of small cells & PW: Decrease with the number of small cells \\
		& and become saturated for large scale case & and become saturated for large scale case\\
\hline
\hline
\end{tabular}
\end{table}
\textcolor{black}{
\subsection{Infrastructure}
Due to the massively connected devices, the actual traffic for 5G networks will not only scale up in volume but also in dynamics. To deal with this effect, the fundamental solution is to introduce the network heterogeneity, where the baseband processing is centralized at the server pool and the radio frequency functionality is distributed over the whole network. Through this approach, the communication tasks will be decoupled into two parts: information processing in baseband and information transportation in radio frequency. In the information processing part, the computational capability and power for baseband processing can scale with the chipset evolution and enjoy the benefits of Moore's Law \cite{JR:Moore98}. In the information communication part, the radio frequency components can be massively deployed to increase the transportation distance, and adopt tailor-made design based on traffic types and the activation/de-activation strategies based on the traffic variation.
}

\textcolor{black}{
\subsection{Protocol}
In the protocol level, the traditional control and user signaling are coupled together in order to control the transmission delay and balance the reliability. With diversified applications of 5G networks, a single solution to cover all throughput, delay, and connectivity requirements is not feasible. Meanwhile, the evolution on software defined property for backhaul networks, such as network function virtualization (NFV) \cite{NFV:12} and software defined network (SDN) \cite{JR:Goth11}, has triggered the radio access networks to merge with the same trend  \cite{JR:Xu13, JR:Liu14} and has been therefore presented as key technologies with great potential in terms of green networking.} \textcolor{black}{However, SDN switching devices may not directly provide benefits in energy reduction in operating networks while SDN could provide significant promises in reducing network-wide energy consumption \cite{JR:xia15}. For example, As a proof,
energy-aware data link adaptation has been demonstrated in \cite{CF:heller2010elastictree} with SDN. Specifically, a mechanism to determine minimum data links and switching devices for a data center network is proposed based on traffic loads and dynamically power down redundant links and switching devices for energy efficient operations.
 A heterogeneous protocol solution with centralized control plane and separated user plane to accommodate heterogeneous wireless applications is thus required. Based on this framework, the mobility management across heterogeneous wireless standards can be performed via a unified control plane to provide seamless user experience \cite{JR:Yazici14} and the user plane protocols/standards can be adjusted based on the heterogeneous requirements and implemented via a software defined environment. }
\textcolor{black}{
\subsection{Implementation}
To cope with the heterogeneous applications in 5G networks, a homogeneous implementation principle only relying on software defined properties may fail to achieve the best EE performance because the dynamic circuit power of the homogeneous hardware system is proportional to the square of clock frequency \cite{BK:Rabaey} and causes overwhelming energy consumption when running on non-urgent applications. Inspired by the HetNet infrastructure and protocol, we propose to use the heterogeneous implementation framework with a cloud-based centralized computing platform and several types of separated plug-in accelerators\footnote{For instance, general purpose graphical processing units (GPGPU), field programmable gate array (FPGA) and application specific integrated circuits (ASIC) can be regarded as different types of accelerators for parallel processing, reconfigurable computing and customized operations respectively.}. This idea takes the advantage of high performance energy efficient computing for most of common applications while tackling the extreme cases using specific hardware structures. Meanwhile, we can utilize the programmable features for part of hardwares to quickly adapt and optimize for changing protocols and applications to avoid unnecessary hardware and energy costs.}

\section{Conclusions}\label{sec:con}
In this article, we have provided a comprehensive survey on the research progress on fundamental green tradeoffs of typical 4G and 5G communication technologies, including OFDM and NOA, regular and massive MIMO, HetNets and UDNs. Theoretical achievements as well as practical energy efficient schemes have been summarized according to the tradeoff framework, followed by some in-depth analysis on the open technical issues. With the explosive demands for high data rate and massive connections, 5G networks will experience a fundamental breakthrough in all related aspects, where we believe the fundamental green tradeoffs shall be evaluated as a key enabler to keep the affordable energy consumption.

\bibliographystyle{IEEEtran}
\bibliography{IEEEabrv,mybib}

\end{document}